\documentclass[11pt, fleqn]{article}

\usepackage{amsmath}
\usepackage{amssymb}
\usepackage{amsfonts}
\usepackage{amsthm}
\usepackage{dsfont}
\usepackage{mathrsfs}
\usepackage{ulem}
\usepackage{graphicx}

\usepackage{mdframed}

\usepackage{enumerate}

\usepackage[arrowdel]{physics}
\usepackage{tensor}

\usepackage{simplewick}
\usepackage{feynmf}

\usepackage{tabu}
\usepackage{physics}
\usepackage{mathdots}

\newcommand{\operator}[1]{\hat{#1}}

\newcommand{\R}{\mathbb{R}}

\newcommand{\Z}{\mathbb{Z}}

\newcommand{\T}{\mathbb{T}}

\newcommand{\pauli}[1]{
    \ifnum#1=1
        \operator{\sigma}_{x}
    \else
        \ifnum#1=2
           \operator{\sigma}_{y}
        \else
            \ifnum#1=3
                \operator{\sigma}_{z}
            \else
                \errmessage{Incorrect number given to pauli}
            \fi
        \fi
    \fi
}

\usepackage{geometry}
\usepackage{comment}

\usepackage{soul}
\usepackage{subcaption}

\usepackage{jheppub}
\usepackage{amsmath,amssymb,amsthm}
\usepackage{bm}
\usepackage{slashed}
\usepackage{graphicx}
\usepackage[T1]{fontenc}
\usepackage{fix-cm}

\usepackage{tabu}
\usepackage{physics}

\makeatletter
\newcommand*\bigcdot{\mathpalette\bigcdot@{.5}}
\newcommand*\bigcdot@[2]{\mathbin{\vcenter{\hbox{\scalebox{#2}{$\m@th#1\bullet$}}}}}
\makeatother

\newlength{\dummysp}
\newcommand{\ba}{\begin{eqnarray}}
\newcommand{\ea}{\end{eqnarray}}

\usepackage[font=scriptsize]{caption}
\usepackage{mwe}

\def\R{{\mathbb R}}
\def\S{{\mathbb S}}
\def\Z{{\mathbb Z}}
\def\T{{\mathbb T}}

\def\tr{\,{\rm tr}\,}

\title{$\mathbf{D}$-branes and fractional  instantons on a twisted $\mathbf{\T^4}$: the moduli space as an $\mathbf{{\cal N}}\mathbf{=2}$ supersymmetric Higgs branch}
 
 \author{Erich Poppitz} 
 \affiliation{Department of Physics,   University of Toronto, 60 St George St., 
Toronto, ON M5S 1A7, Canada}
\emailAdd{erich.poppitz@utoronto.ca}   
 
 \abstract{
  \smallskip
 
{\flushleft{We}} study self-dual instantons of topological charge $Q$$=$${r\over N}$,  $\forall r$$\in$$\mathbb{N}$,  in $SU(N)$ Yang-Mills theory on  $\T^4$ with 
 't Hooft twists, by embedding them into worldvolume theories of $D$-branes. 
 To study their moduli, we construct the wrapped intersecting brane configurations dual to general constant field strength instanton backgrounds. We show that, locally, the moduli space  is identified with the Higgs branch of an ${\cal N}$$=$$2$ supersymmetric theory. This parameterization of the moduli space is equivalent to one recently found in field theory, but is obtained with significantly less effort and  has manifest hyper-K\" ahler structure. 
Our hope is that  combining different perspectives on instantons on the twisted torus will  help understand the still unknown global structure of the moduli space for general solutions with $Q$$=$${r\over N}$ as well as the nature of instantons with all moduli turned on---when some $Q$$<$$1$ and all $Q$$\ge$$1$  instantons become space-time dependent. For integer $Q$, these are expected to match the ADHM solution in an appropriately taken infinite volume limit.}

\begin{document}

\maketitle
 
\section{Introduction, summary, and outlook}
\subsection{Motivation}

That instantons of topological charge $r\over N$ in $SU(N)$ gauge theories exist, on $\T^4$ with twisted boundary conditions, was shown by 't Hooft in his work on quark confinement  \cite{tHooft:1979rtg,tHooft:1981sps}. He also found a class of explicit solutions, the constant field strength\footnote{Throughout, we call these the ``constant-$F$'' solutions; their gauge invariant property is that all invariants constructed from the field strength $F$ are space-time independent.}  fractional instantons \cite{tHooft:1981nnx}. These are self-dual (BPS, or minimal action) for appropriately tuned ratio of periods of $\T^4$, or simply, for a tuned shape of the torus. Their study is the main subject  of this paper.

\subsubsection{A brief history of fractional instantons and their semiclassical uses}
\label{sec:history}

To motivate our interest in fractional instantons from a physics point of view, we begin with a lightning review of their semiclassical role  in the nonperturbative dynamics of gauge theories. Our intention is not  to list the numerous interesting  older and recent papers on the subject, but only to give, to the interested reader, a sense of the what has been achieved in this area of research over the years.
 For  an overview of the scope and results of this paper, one can  jump directly to Section \ref{sec:summary}
 
For many years, the study of fractional instantons and the associated nonperturbative dynamics of gauge theories was pursued by only a few---notably,  the Madrid group, whose many contributions were recently  reviewed by Gonz\' alez-Arroyo  \cite{Gonzalez-Arroyo:2023kqv}. A highlight is their work from the 1990s  \cite{RTN:1993ilw, Gonzalez-Arroyo:1995ynx}, which showed that fractional instantons are responsible for semiclassical\footnote{Semiclassics holds when the $\T^3$ size $L$ is small, $L  N \Lambda \ll 1$, where $\Lambda$ is the gauge theory strong coupling scale; this remark also applies to the applicability of semiclassics on $\R^3 \times \S^1$ or $\R^2 \times \T^2$.}  confinement on $\R \times \T^3$ and  numerically demonstrated its ``adiabatic continuity'' to confinement in the strongly-coupled $\R^4$-limit. Closer to the subject of this paper, since the early 2000s, they also developed an  analytical approach to the construction of approximate non-constant fractional instanton solutions  \cite{GarciaPerez:2000aiw,Gonzalez-Arroyo:2019wpu}, obtained from  't Hooft's by slightly deforming the torus  away from the self-dual shape. 

Somewhat later, a seemingly unrelated---at the time (see the last paragraph in this Section)---development  was inspired by the work of \" Unsal from 2007 \cite{Unsal:2007jx,Unsal:2007vu}. He showed that objects of fractional topological charge were behind    semiclassical  confinement and chiral symmetry breaking on $\R^3 \times \S^1$. The fractionally-charged objects are  the so-called ``monopole-instantons;'' see the review \cite{Poppitz:2021cxe} for an extensive  list of references.

The more recent interest in the subject was driven by the improved understanding of generalized symmetries  and especially of their anomalies, emerging after \cite{Gaiotto:2014kfa,Gaiotto:2017yup}. The connection to the old picture is that 't Hooft twists \cite{tHooft:1979rtg} are now seen as a topological background of the $2$-form $\Z_N$ gauge field gauging the $\Z_N^{(1)}$ $1$-form center symmetry; this interpretation helps identify  various 't Hooft anomalies involving $\Z_N^{(1)}$.

In particular, the understanding of anomalies of generalized symmetries    in the language of  \cite{Cox:2021vsa} allowed Mohamed Anber and the author   
to complete the semiclassical calculation of the (higher-order) gaugino condensate in super-Yang-Mills theory  via fractional instantons on $\T^4$ \cite{Anber:2022qsz,Anber:2023sjn,Anber:2024mco,Anber:2025vjo}. After understanding many subtleties, we obtained a result in agreement with the $\R^4$ calculations of higher-order gaugino condensates  \cite{Dorey:2002ik}, which used the ADHM construction \cite{Atiyah:1978ri,Atiyah:1979iu}. The present work stems from the desire to better understand the properties of fractional instantons noted in these works and in \cite{Anber:2024uwl,Anber:2025yub}, as  reviewed in Section \ref{sec:why} below.

To end this short overview, we should mention another recent development  involving fractional topological charge objects. Semiclassical ideas of confinement  were also applied to $\R^2 \times \T^2$ with 't Hooft twists 
\cite{Tanizaki:2022ngt}, where it was shown that ``center vortices,'' another kind of objects of fractional topological charge are responsible for confinement (earlier, this had been mentioned in \cite{Gonzalez-Arroyo:1998hjb,Montero:2000pb}, also studied in their recent   \cite{Bergner:2025qsm}). A related development of the last two years is the realization that the ``center-vortex'' mechanism on $\R^2 \times \T^2$ and the ``monopole-instanton'' one on $\R^3 \times \S^1$ are continuously connected;\footnote{The role of center vortices and   monopoles in confinement, as well   the  relation between them, has long been a subject of (non-semiclassical) lattice studies, see Greensite's monograph \cite{Greensite:2011zz}.}  remarkably, it was shown that, in some limits, this connection can even be explored analytically \cite{Hayashi:2024yjc,Guvendik:2024umd}. The continuity between semiclassical configurations  was also studied using numerical minimization of  lattice actions \cite{Wandler:2024hsq,Dobozy:2026edc}.  It thus appears that all   fractional-charge objects responsible for confinement are ``unified,'' in that they can be obtained from ones on $\T^4$  by changing the twists and taking different limits of the torus periods.

Having described  the dynamical role of these semiclassical objects (noting in passing that very few analytical solutions are known), we now turn  to our main subject. It can be thought of as one of kinematics: we explore  the nature and moduli of general charge $r/N$ self-dual instantons, properties   far from  being thoroughly understood.

\subsubsection{Why study constant field strength BPS instantons on $\T^4$?}
\label{sec:why}
This paper studies the constant-$F$ instantons on $\T^4$ with 't Hooft twisted boundary conditions. These solutions are BPS when a ratio of the $\T^4$ periods is appropriately tuned, e.g. the torus shape is fixed.\footnote{For the impatient, see eqn.~(\ref{tuneintro}).} The minimal action (BPS, or self-dual) solutions on    $\T^4$ with  a  shape thus fixed are the object of  interest here.  

One might ask why such solutions constant over all of space-time are of interest? After all, in the large-$\T^4$ limit, their constant action density spreads out thinly  and is completely delocalized, making one wonder how they can be relevant to  local  dynamics.
Our answers to this question are:
\begin{enumerate}
\item If the $\T^4$ shape is slightly detuned  from the BPS value, with the detuning measured by a small dimensionless $\Delta$, approximate space-time dependent analytic solutions of minimal action can be obtained from the constant-$F$ ones by means of the ``$\Delta$-expansion''   \cite{GarciaPerez:2000aiw,Gonzalez-Arroyo:2019wpu}. These approximate solutions,  generalized in   \cite{Anber:2023sjn} to include multi-fractional instantons, were an indispensable tool in the calculation of gaugino condensates. 
Thus, one reason that understanding the constant-$F$ BPS backgrounds and their moduli (or self-dual deformations) is important is because they are the first step in the construction of approximate  self-dual solutions on the detuned $\T^4$. 
\item It turns out that even the constant-$F$ solutions for tuned $\T^4$ shape, i.e.~at $\Delta=0$, have more interesting properties than   meets the eye. A puzzling feature was  noted in \cite{Anber:2023sjn}, explained later in \cite{Anber:2025yub}:  many of the constant-$F$ BPS solutions have extra (we shall call them ``missing'') moduli.\footnote{The fluctuations around some constant-$F$ BPS solutions on $\T^4$ were studied early on by van Baal \cite{vanBaal:1984ar}, but  these were not sufficiently general to have ``missing'' moduli.} When turned on, they  make the tuned-$\T^4$ solutions space-time dependent. In other words, the constant-$F$ solutions are a set of measure zero in the self-dual  moduli space.
Via a local analysis of the moduli space, we showed that such extra moduli  exist for some of the   solutions with $Q < 1$ (i.e. $1<r<N$) and for all solutions with $Q \ge 1$ (i.e. $r \ge N$). In particular, all integer-$Q$ solutions on the fixed-shape $\T^4$ are, for generic values of the moduli, space-time dependent. It is natural to expect  that they approach the ADHM one when their overall size is smaller than that of $\T^4$; how precisely this limit is achieved is an interesting open question.

The global structure of the moduli space of the general  BPS solutions with $Q={r/N}$ on the tuned-$\T^4$ is currently not understood. On the one hand, it would help elucidate how finite volume and infinite volume solutions are related. From a calculational point of view, self-dual-only backgrounds are  useful 
for calculating protected (holomorphic) quantities in supersymmetric theories.\footnote{See, however, the recent \cite{Unsal:2026aax}.} In this regard, understanding the general moduli space would help finish the program of \cite{Anber:2022qsz,Anber:2023sjn,Anber:2024mco,Anber:2025vjo} by calculating the $r$-point gaugino condensate\footnote{The calculation in these references applies to $r < N$ only.} for all values of $r$. It might also be of interest from a mathematical point of view, where the only related result we are aware of is \cite{BraamTodorov:1992}.
\end{enumerate}
In \cite{Anber:2025yub}, we performed a local linearized analysis of the self-dual perturbations around the constant-$F$ solutions. Using the rather laborious calculations of \cite{Anber:2023sjn}, we obtained a local parameterization of the moduli space, which led to the results on space-time dependence stated in point $2.$ above.
To make further progress, 
we believe that it is  of interest to develop alternative points of view on these, fractional or not, constant-$F$ BPS instantons and on their moduli. 
Our hope is that this will  help improve their understanding and aid future studies.

\subsection{Summary}
\label{sec:summary}
In Section \ref{sec:wordy}, we summarize the main points of our study and the results, skipping most technical details. Then, in Section \ref{sec:concrete}, we give a picture and discussion of one simple example   illustrating our findings. A reader not interested in all details  should find this summary Section  sufficient to get an idea about our results.
\subsubsection{A wordy description of the approach and results}
\label{sec:wordy}

The point of view we take in this paper is that $D$-branes in type II string theory are the natural setup,\footnote{Here we essentially  quote  Tong's lectures \cite{Tong:2005un}, which contain a concise introduction to the topic.}
 from a physicist's point of view, to understand general self-dual instantons. The brane picture is particularly useful to study  the ADHM \cite{Atiyah:1978ri,Atiyah:1979iu} multi-instanton\footnote{Incidentally, the  self-dual ``monopole-instantons'' on $\R^3 \times \S^1$ (mentioned in Section \ref{sec:history}) were also independently discovered using $D$-branes \cite{Lee:1997vp}.} moduli space and background \cite{Witten:1994tz,Douglas:1995bn,Douglas:1996uz}.

Thus inspired, we begin by embedding 't Hooft's constant-$F$ self-dual configurations with $Q_{SU(N)}={r/N}$ into the worldvolume   theory of $N$ $D_{p+4}$ branes wrapped on $\T^4$. The embedding into a  $U(N)$ bundle requires turning on appropriate $U(1)$ fluxes, or first Chern characters, in appropriate two-planes of $\T^4$. These steps,    involving only QFT tools,  taken from \cite{Anber:2024uwl}, are described in detail in Section \ref{qftsetup}.  The $\T^4$ shape is appropriately tuned  such that these  backgrounds are BPS---breaking one-half the supersymmetry, thus preserving $8$ supercharges. 

For later comparison with the $D$-brane study, we quickly review, in Section \ref{sec:missing},
 the 
  ``missing moduli'' puzzle  and the result of its  resolution in QFT \cite{Anber:2025yub} (the associated extra moduli were already  mentioned  in point $2.$ of Section \ref{sec:why}).

To study self-dual   fluctuations 
  using world-sheet tools and to explore the moduli space in the brane setup, we perform   $T$-duality  in two of the $\T^4$ directions; this is useful since $T$-duality removes the flux,  mapping it into tilted branes (see e.g.~\cite{Polchinski:1996fm}). To $T$-dualize, in  Section \ref{sec:convenient} we choose a convenient gauge \cite{Taylor:1996ik,Taylor:1997dy} for the transition functions. To arrive at the dual brane configuration,\footnote{For a study of the BPS conditions, as in  \cite{Berkooz:1996km}, on the shape of the dual $\tilde{\T}^4$ in the intersecting brane picture, see Figure \ref{brane4} on the covering space and Section \ref{sec:bps}. As expected,  the results are identical to  the QFT ones of Section \ref{sec:constant}.} a careful interpretation of the  transition functions is required, described in Sections \ref{sec:branesdual} and \ref{sec:numberofintersect}. The end result is that the  $T$-dual configuration consists of two stacks of  $D_{p+2}$ branes wrapped  on intersecting two-cycles of the dual $\tilde{\T}^4$. The final brane picture, the determination of the number of  intersections of the wrapped stacks of $D_{p+2}$ branes, and its consistency   with the various Ramond-Ramond (RR) charges and $U(1)$ fluxes, is discussed in Section \ref{sec:numberofintersect}.\footnote{Earlier studies of brane configurations $T$-dual to constant-$F$ backgrounds, including some of  't Hooft's, are \cite{Guralnik:1997sy,Hashimoto:1997gm}. However, the instanton moduli space was not their focus;  also, the backgrounds considered were not general enough to exhibit ``missing'' moduli (i.e. have more than one  intersection point, see below).}

The result for the brane configuration dual to the general 't Hooft solutions we consider  is pictorially shown on Figure \ref{brane6} in Section \ref{sec:numberofintersect} and described in its caption. In the following Section \ref{sec:concrete} of this Introduction, we give a simple concrete example of such a configuration illustrating the main points, see Figure \ref{brane5}. 
 The number of  intersections between the two stacks depends on the  instanton considered and is crucial for the study of the moduli space. The reason is that   each  intersection supports localized massless string excitations.\footnote{In the string phenomenology literature,  nonsupersymmetric  constructions   of wrapped   branes with multiple  intersections have been used to obtain   Standard Model-like setups, with chiral fermions of different generations  localized at the different  intersection points, see e.g.~\cite{Aldazabal:2000cn,Anastasopoulos:2011kr}.} With 8 supercharges preserved, these form a hypermultiplet, bifundamental under the gauge groups on the two intersecting branes  \cite{Berkooz:1996km}; we  show that these extra massless modes, whenever there is more than a single  intersection, contain the ``missing moduli.'' 
  
  To study the moduli space locally, using intuition from ${\cal{N}}=2$ supersymmetry in 4d, we choose $p=3$. As in the presentation of the ADHM moduli space in \cite{Tong:2005un}, this makes the noncompact part of the $D$-brane worldvolume four dimensional. Ignoring  $\tilde{\T}^4$-variations of the fields (including the compactness of brane-position moduli, as well as the fact that different hypermultiplets are localized at different points on $\tilde{\T}^4$), the long distance theory on the noncompact part of the volume of  the  two stacks of $D_5$ branes (each wrapped on some two-cycle of the torus) is an 8-supercharge 4d theory with a product gauge group, one gauge group on each stack. 
  In ${\cal N}=2$ theories, the superpotential is completely determined by supersymmetry and the matter content. Thus, the study of the moduli space reduces to   parameterizing the conditions for vanishing of the $D$- and $F$-terms along the Higgs branch of the ${\cal N}=2$ worldvolume theory.\footnote{On the Coulomb branch, the two stacks are separated in directions orthogonal to $\tilde{\T}^4$, see Section \ref{sec:concrete}.}  This study of the moduli space is local, since  the 4d EFT does not capture the compact nature of some of the fields, such as brane positions in $\tilde{\T}^4$, as well as any $\tilde{\T}^4$-variations, which should become important when some of the ``missing moduli'' are turned on. 
  
  Nonetheless, our point is that this 4d ${\cal{N}}=2$ EFT reproduces the results for the moduli-space parameterization of \cite{Anber:2025yub}
in  a few lines, essentially without any calculation.  Those interested in the  analysis of the ${{\cal N}}=2$ Higgs branch for general solutions are invited to consult Sections \ref{sec:u1branch} and \ref{sec:enhanced}. 

\subsubsection{A concrete example   illustrating the brane picture and moduli space}

\label{sec:concrete}
{\flushleft{Here, we discuss an example with the minimum detail required to illustrate the results.}}

{\flushleft\bf{Minimum QFT background:}} Without much ado,  we consider the following  $SU(N)$    backgrounds on  $\T^4$. They depend on $N$ and two positive integers, $k$ and $r$, with $N = k + \ell$ ($\ell >0$):
\begin{eqnarray}\label{SUNbackgroundIntro}
{ F}_{12} &=&\left(\begin{array}{cc} - {2 \pi \ell r   \over N k L_1 L_2} {I}_k & 0 \cr 0 & {2 \pi r  \over N L_1 L_2} {I}_\ell \end{array}\right), ~~{ F}_{34} = \left(\begin{array}{cc} - {2 \pi   \over N L_3 L_4} {I}_k & 0 \cr 0  & {2 \pi k \over N \ell L_3 L_4} {I}_\ell \end{array}\right)~.
\end{eqnarray}
 We use $L_\mu$, $\mu=1,...4$, to denote the rectangular $\T^4$ periods. $I_k$ and $I_\ell$ are  $k \times k$ and $\ell \times \ell$ unit matrices. 
  Clearly, (\ref{SUNbackgroundIntro}) is self-dual provided $F_{12} = F_{34}$, i.e. 
 \begin{eqnarray}\label{tuneintro}{L_1 L_2 \over L_3 L_4} = {  \ell r \over k }. \end{eqnarray} We assume from now on that the shape of the rectangular $\T^4$ is thus tuned.
The  topological charge is 
\begin{eqnarray}\label{chargeintro}
Q_{SU(N)} = {r\over N},~ r \in \mathbb{N},
\end{eqnarray} as  a simple calculation using (\ref{SUNbackgroundIntro}) shows. The $SU(N)$ background $F$ is the traceless part of the  $U(N)$ background $\cal{F}$, i.e.~$F = {\cal F} - {I_N \over N} \text{tr}\; {\cal F}$: 
\begin{eqnarray}\label{UNbackgroundIntro}
{\cal F}_{12} &=&\left(\begin{array}{cc} - {2 \pi  r  \over k L_1 L_2} {I}_k & 0 \cr 0 &0 \times I_\ell\end{array}\right), ~~
{\cal F}_{34} =\left(\begin{array}{cc}0\times I_k & 0 \cr 0 & {2 \pi  \over \ell L_3 L_4} {I}_\ell \end{array}\right).
\end{eqnarray}
In other words, the $U(N)$ background is obtained from the $SU(N)$ one by adding appropriate $U(1)$ fluxes in the $12$ and $34$ planes.\footnote{Here, we have only added the minimal $U(1)$ fluxes such that the $U(N)$ topological charge of (\ref{UNbackgroundIntro}) vanishes.} The values of the first Chern characters ($U(1)$ fluxes) and the second Chern character ($U(N)$ topological charge) determine the $SU(N)$ topological charge (\ref{chargeintro}). See Section \ref{sec11} for explicit formulae.

The $U(N)$ background (\ref{UNbackgroundIntro}), as opposed to (\ref{SUNbackgroundIntro}), is obviously not self dual, but it does not break supersymmetry if (\ref{tuneintro}) holds (see Section \ref{sec:bps} for discussion of supersymmetry). In Section \ref{qftsetup}, we introduce other  details, too bulky to show here: the $\T^4$ transition functions and the moduli.\footnote{We stress that the simplicity of the background (\ref{UNbackgroundIntro}) is deceiving: its transition functions are quite involved and the  details are crucial for obtaining the $T$-dual picture.} 
From these, what matters most  for our short presentation here is that  there are $4 \text{gcd}(k,r) + 4$ constant moduli (holonomies, or Wilson lines) that commute with the transition functions associated with (\ref{UNbackgroundIntro}). Four of these are associated with the $U(1)$ Wilson lines, one per each direction of $\T^4$. On the other hand, the $SU(N)$ transition functions only allow $4 \text{gcd}(k,r)$ Wilson lines. This is the origin of the ``missing moduli'' problem. It arises because the index theorem \cite{Schwarz:1977az,Weinberg:1979ma,Taubes:1982qem} (which determines the dimension of the moduli space for self-dual configurations in $SU(N)$ with topological charge $r/N$) demands the existence of $4 r$ moduli. Thus, if gcd$(k,r) \ne r$, extra moduli are needed. These $4 r - 4 \text{gcd}(k,r)$ moduli were  found in the study of linearized self-dual perturbations \cite{Anber:2025yub}, benefitting from the long-winded calculations of \cite{Anber:2023sjn}. The results  of that reference are described in Section \ref{sec:missing}, for easy later comparison with the $D$-brane results.

{\flushleft\bf{D-brane realization and an example:}} We now imagine that $x_1,...,x_4$, the $\T^4$ coordinates, are part of the worldvolume of $N$ $D_{p+4}$ branes, where  $p+1$ of the worldvolume coordinates are extended (these are $x_0, x_{4+1},...x_{4+p}$). We perform $T$-duality  in $x_2, x_4$, a choice determined by the transition functions. We then obtain a configuration on the dual   $\tilde{\T}^4$. The $D_{p+4}$ branes with worldvolume flux (\ref{UNbackgroundIntro}) give rise to two stacks of $D_{p+2}$ branes wrapped on appropriate $2$-cycles of $\tilde{\T}^4$. We label the dual space coordinates $y_0,..., y_9$, where $y_1,...,y_4$ parameterize the $\tilde{\T}^4$ of dual periods $\hat L_\mu$. The noncompact directions of the worldvolume of the $D_{p+2}$ branes are $y_0, y_{4+1},..., y_{4+p}$. The $D_{p+2}$ branes,   wrapped on two cycles in $\tilde{\T}^4$, are localized in $y_{4+p+1},..., y_9$ and in the two directions of $\tilde{\T}^4$ orthogonal to the cycle they are wrapped on.

Notably, for $p=3$, the $D_5$ branes' noncompact worldvolume is 4d, $y_{0,5,6,7}$, and the $D_5$ branes are localized in $y_{8,9}$.
To give a flavour of the resulting brane configuration, we now consider a particular example. We consider an $SU(8)$ gauge group,  taking $k=6$ ($\ell =2$) and $r=4$, corresponding  to a charge-$1/2$ instanton. Here, gcd$(k,r)=2 \ne r = 4$, hence this is an example with ``missing moduli.''   
 
The configuration on the dual $\tilde{\T}^4$ is that of two stacks of branes. A careful study of the transition functions reveals that the two stacks are as follows.  One  stack consists of gcd$(k,r)= 2$ parallel $D_{5}$ branes (whose mutual separation is a modulus) wrapped on one two-cycle; we call this the ``$k$''-stack. The other stack is one of a single $D_{5}$ brane, we call it the ``$\ell$'' stack, wrapped on a different cycle. The angles between the cycles are such that when the shape of the torus is tuned as in (\ref{tuneintro}), $8$ supersymmetries are preserved. 

\begin{figure}[h] 
   \centering
   \includegraphics[width=6in]{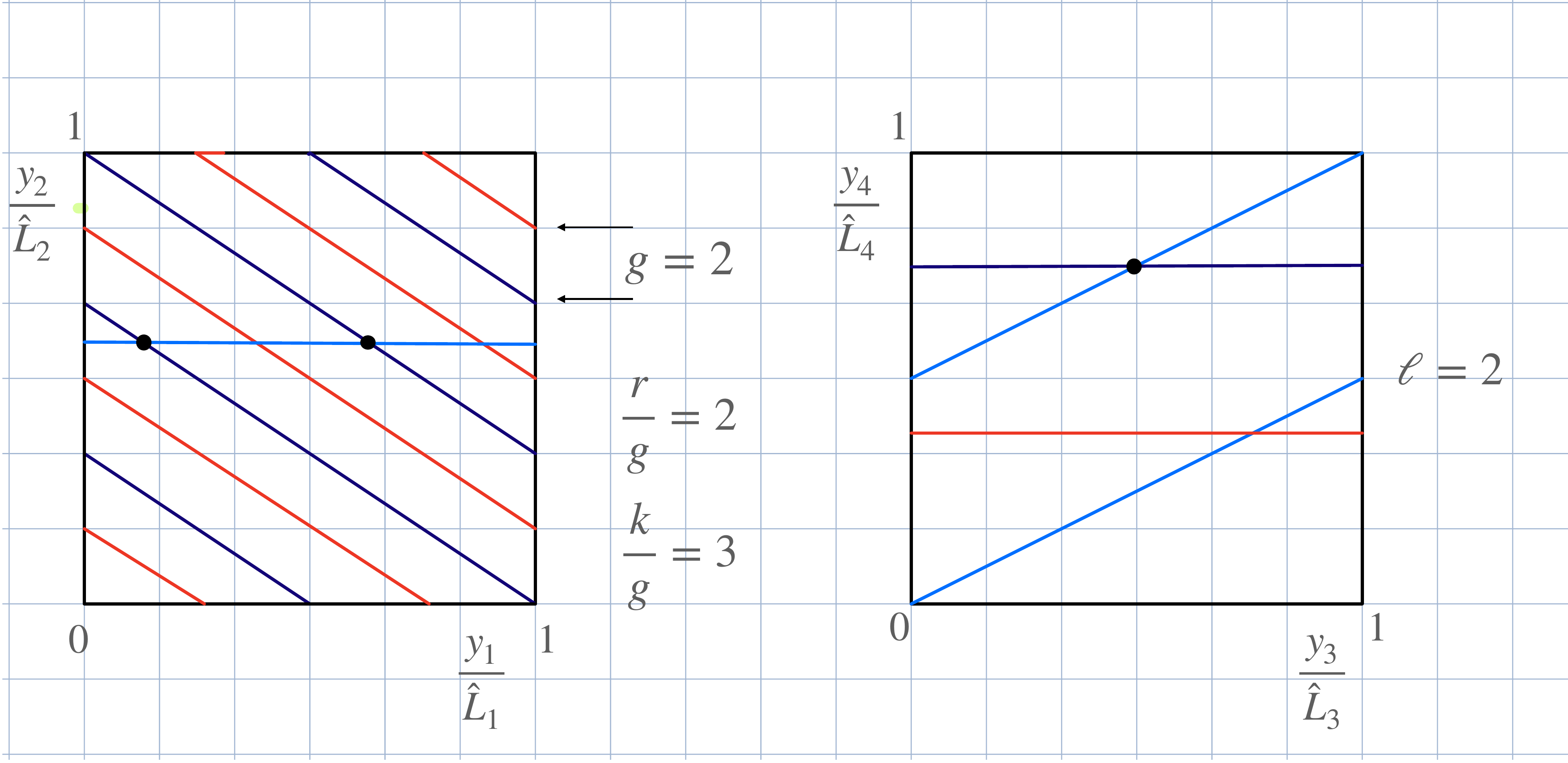} 
   \caption{The two stacks of ``$k$''- and ``$\ell$ ''- $D$-branes wrapped on two-cycles in $\tilde\T^4$,    are shown  in the $12$ and $34$ planes,   for our $SU(8)$ example with $k=6$, $r=4$, $\ell = 2$; $g\equiv \text{gcd}(k,r)=2$. The grid background is kept to help visualize the windings of the branes.  The ``$k$'' stack consists of $g=2$ branes. Its windings, as one dimensional curves in each two-plane,   are ${k \over g}=3$ in $y_1$, ${r\over g}=2$ in $y_2$, unity in $y_3$ and zero in $y_4$, as  pictured.
  The second stack, the single ``$\ell$''-brane,  is shown in light blue. It is parallel to $y_1$, wrapping once around it, but is tilted in the $34$ plane, wrapping $\ell$ times in  $y_3$ and once in $y_4$. We only show, using black dots, the $r/g=2$  intersection points between the dark blue and light blue branes. }
   \label{brane5}
\end{figure}

These cycles intersect on the $\tilde{\T}^4$, and the whole picture is  on Figure \ref{brane5}, where we show the $12$ and $34$ planes of the torus,  plotting $\tilde{\T}^4$ coordinates divided by the corresponding dual period $\hat L_\mu$. The two parallel branes of the ``$k$''-stack are shown in red and dark blue, while the single brane of the ``$\ell$'' stack is shown in light blue.\footnote{These cycles are described by eqns.~(\ref{hatYk}, \ref{hatYell}) with $q_1=q_3=0$. The two stacks in an $SU(N (=8))$ theory, one of gcd$(k,r) (=2)$ branes and another of a single brane, with the windings  shown,  can be seen, Section \ref{sec:numberofintersect}, to have all the correct RR charges.}      
We stress that the $4\text{gcd}(k,r) + 4 = 12$ moduli allowed by the transition functions have clear geometric interpretation: they correspond to the positions and Wilson lines of each brane on $\tilde{\T}^4$ (there are two positions and two Wilson lines for each of the dark blue, red, and light blue brane, giving $3 \times 4 =12$). The ``missing'' moduli come from hypermultiplets localized at the  intersection.

Each of the gcd$(k,r) (=2)$ parallel ``$k$''-branes has $r/\text{gcd}(k,r) (=2)$  intersection points with the light blue ``$\ell$''-brane; these are shown by black dots for the dark blue brane only. At each  intersection point between two branes, there are massless string excitations, a bifundamental hypermultiplet charged under   the  $U(1) \times U(1)$ gauge fields on the worldvolumes of the two intersecting branes.

Following the logic outlined earlier, let  us now  describe the 4d EFT in the noncompact  worldvolume directions of the $D_5$ brane.\footnote{An  introduction to the use of    branes to study  field theories is in \cite{Giveon:1998sr}.}  It is an ${\cal N}=2$ 4d supersymmetric theory with gauge group $U(1)_{\ell} \times U(1)_1\times U(1)_2$, where   $U(1)_\ell$ lives on the light blue brane and the other two $U(1)$ factors on  dark blue and red branes. 
Each $U(1)$ factor has an adjoint chiral superfield (in ${\cal{N}}=1$ notation) in the ${\cal{N}}=2$ vector multiplet, which describes the position of the brane in the $y_{8,9}$ direction. There is also an ${\cal{N}}=2$ adjoint hypermultiplet, whose two chiral adjoint supermultiplets (also using ${\cal{N}}=1$ notation) describe its position in $\tilde{\T}^2$ and the two Wilson lines in the compact worldvolume directions. Finally, there are 
$r/{\rm gcd}(k,r) = 2$ massless hypermultiplets coming from the  intersection points between the dark blue (say, labelled by $i=1$) and red ($i=2$) branes  with the light blue ``$\ell$'' brane. These hypers have charges $(1,-1)$ under $U(1)_\ell \times U(1)_i$. 

The part of the moduli space of the ${\cal{N}}=2$ EFT that has the interpretation of the instanton moduli space is identified with the Higgs branch of the theory. This is because the Coulomb branch (giving vevs to the adjoint scalar in the vector multiplet) corresponds to separating the branes along the $y_{8,9}$ noncompact directions. Such a configuration, however,  has nothing to do with the original $D_{p+4}$ configuration in the $U(N)$ theory with flux (it would correspond to starting on the Coulomb branch of the original  $16$-supercharge theory). 

We shall not present the analysis of the Higgs-branch moduli space in this introductory Section. It is well known that ${\cal N}=2$ supersymmetry completely determines it, given the hypermultiplet matter content. We   refer to Section \ref{sec:u1branch} for a study of the Higgs branch, and the derivation of eqns.~(\ref{fterm2}, \ref{dterm2}) and (\ref{dimensionmoduli}), also reproduced below in (\ref{fterm2intro}, \ref{dimensionmoduliintro}).

We do, however, give the resulting description of the Higgs branch for  general $N, r, k$  as well as the counting of the moduli. Denoting $g \equiv \text{gcd}(k,r)$, 
there are the $4 g +4$ moduli corresponding to brane positions/Wilson lines, or adjoint hypers, that we already mentioned. In addition, there are 
the $r/g$ hypermultiplets, living on the $r/g$  intersections between each of the $g$ branes  of the ``$k$'' stack (indexed by $i=1,...,g$) and the single brane of the ``$\ell$'' stack. Their bosonic components are the complex scalars $q_i^a, \tilde{q}_a^i$, $a=1,...,r/g$, each charged under $U(1)_\ell \times U(1)_i$ (the gauge group, for generic positions of the moduli, is $U(1)_\ell \times U(1)^g$).   These hypermultiplets comprise $4\times  {r \over g} \times r = 4 r$ real variables.
These variables obey the $F$- and $D$-term Higgs-branch conditions \begin{eqnarray}\label{fterm2intro}
\sum_{a=1}^{r\over g} \tilde{q}_a^i q_i^a &=& 0, ~\forall \; i=1,...g, ~~\text{$2 g$ real constraints},\nonumber\\
\sum_{a=1}^{r\over g} |\tilde{q}_{a}^i|^2 - |q_{i}^a|^2 &=& 0,  ~\forall \; i=1,...g,~~\text{$g$ real constraints}.
\end{eqnarray}
One immediate observation is that when gcd$(k,r)=r$, i.e. $r/g=1$, the only solution of these equations is $q_i = \tilde{q}^i=0$, consistent with the absence of ``missing'' moduli in this case.

More generally, adding the number of fields and subtracting the number of constraints, including the modding out by $U(1)_i$, $i=1,...,g$ gauge  transformations,\footnote{Modding out by $U(1)_\ell$ is redundant, see Section \ref{sec:u1branch}.} we find that the real dimension of the Higgs branch is:
\begin{eqnarray}\label{dimensionmoduliintro}
\text{dim(Higgs)} = \underbrace{4 g + 4}_{\text{adjoint hypers}} + \underbrace{4r}_{\text{bifundamental hypers}} - \underbrace{4 g}_{D-, F-, {\text{gauge constraints}}} = 4r + 4.\end{eqnarray} 
 The above result for the dimension, as well as equations (\ref{fterm2intro}), exactly agree with    the index theorem and the moduli space description  of \cite{Anber:2025yub}. This is a very satisfying result, at least to the author. To get a sense why, the reader can compare with the calculations\footnote{In particular, compare the calculations in the voluminous Appendices A and B of \cite{Anber:2023sjn},  with the ones   in Section \ref{sec:u1branch}  leading to (\ref{fterm2intro}, \ref{dimensionmoduliintro}). The latter would be standard fare for an introductory  class on supersymmetry.} that led us to obtain eqns.~(\ref{fterm2intro}, \ref{dimensionmoduliintro}) in QFT (these equations are reproduced in Section \ref{sec:missing}, see eqn.~(\ref{constraintssubbed}), using the notation of \cite{Anber:2025yub}).
 
Finally, we note that at special values of the moduli  there is an enhancement $U(1)^g \rightarrow U(g)$, and the $r/g$ hypers become bifundamentals under $U(g) \times U(1)_\ell$  (on Figure \ref{brane5}, this happens when the red and dark blue parallel branes overlap). As shown in Section \ref{sec:enhanced}, as  usual in ${\cal N}=2$ theories, the Higgs branch conditions can  be cast in an  $SU(2)_R$ invariant form 
\begin{eqnarray}\label{hyperkintro}
\sum\limits_{a=1}^{r \over g}   \chi_a^{\dagger \;i} \sigma^C \chi^a_j -  i \eta_{\mu\nu}^C\;    [X_\mu, X_{\nu}]_j^{\;\; i} &=&0, ~  X_\mu^\dagger = X_\mu, ~i,j = 1,...,g, \;g \equiv \text{gcd}(k,r).
  \end{eqnarray}
The fields $\chi_j^a$  are $r/g$  $SU(2)_R$ doublets, the bifundamental hypermultiplets\footnote{Whose $SU(2)_R$ indices are the only ones not shown; $\sigma^C$ are Pauli matrices acting on the $SU(2)_R$ indices and $\eta_{\mu\nu}^C$ are 't Hooft symbols.}  which are also $U(g)$ fundamentals (indexed by $i,j=1,...,g$) and $(X_{\mu})_j^{\;\; i}$ are $U(g)$ hermitean adjoint fields. The dimension of the moduli space (\ref{hyperkintro}) modded by $U(g)$ gauge transforms, naturally, is the same as (\ref{dimensionmoduliintro}). For  diagonal $X_\mu$, breaking $U(g) \rightarrow U(1)^g$,  (\ref{hyperkintro}) reduces to (\ref{fterm2intro}). While the $SU(2)_R$ invariant form (\ref{hyperkintro}) is superficially similar to the ADHM  moduli space,\footnote{The ADHM moduli space of charge-$q$ $SU(N)$ instantons on $\R^4$ is the same eqn.~(\ref{hyperkintro}), but valued  in the adjoint of $U(q)$ and with $N$ $SU(2)_R$ doublets instead (i.e. with $i,j = 1,...,q$, and $a =1,...,N$). For $q=1$, $X_\mu$ are the position moduli and the $N$ $SU(2)_R$ doublets $\chi^a$, obeying the constraint, include the size, etc., moduli.} we stress that it is only a local description of the $\T^4$-instanton moduli space and that the relation between the two remains to be understood.
  
{\flushleft{Finally,}} with a view towards discussing the outstanding issues, we now summarize what is known and what is not, from this and previous work, about the moduli space of the constant-$F$ solutions (\ref{SUNbackgroundIntro}) on the tuned twisted $\T^4$  parameterized by $(N, r, k)$:
  \begin{eqnarray}\nonumber
  \begin{array}{c||c|c|c} 
  Q=r/N, r \in \mathbb{N}& r< N, \text{gcd}(k,r)=r &r < N, \text{gcd}(k,r) \ne r & r \ge N \cr 
  \hline
  \hline
  4r \; \text{moduli} &4 r  \; \text{adj.~hypers} &4 \text{gcd}(k,r)\; \text{adj.~hypers} +  q, \tilde{q} & 4 \text{gcd}(k,r) \; \text{adj.~hypers} + q, \tilde{q} \cr    &  q = \tilde{q}=0&\text{ small} \; q, \tilde{q} \; \text{obey}\; (\ref{fterm2intro}) &\text{ small} \; q, \tilde{q} \; \text{obey}\; (\ref{fterm2intro}) \cr
    \hline
  x_\mu\text{-dependent?}&{\text{no, for all moduli}}& {\text{yes, provided}} \; q, \tilde{q} \ne 0,& {\text{yes, provided}} \; q, \tilde{q} \ne 0\cr
  \hline
    \text{global structure}& (\prod\limits_{\mu=1}^4 {(S^1)^r \over Z_r})/S_r&\text{?}& \text{?} \cr
    &\text{compact, nonsingular~\cite{Anber:2024mco} }&& \text{for} \; r=N=2, \text{see~\cite{BraamTodorov:1992}}
    \end{array}
  \end{eqnarray}
  
The first column refers to $r <N$ solutions with gcd$(k,r)=r$, for which  \cite{Anber:2025yub}  showed that the solutions are constant throughout the moduli space, a result consistent with the Higgs branch equations (\ref{fterm2intro}) of this paper (which demand $q=\tilde{q}=0$ for this case). The global structure of the moduli space  determined in \cite{Anber:2024mco} shows that the moduli space is smooth and nonsingular (here, we only stress that the various identifications shown act freely; see Section \ref{sec:missing} for more comments on this).

  The second column refers still to $Q < 1$ solutions albeit with gcd$(k,r)\ne r$, for which the Higgs branch eqns.~(\ref{fterm2intro}) of this paper were previously derived in field theory  \cite{Anber:2025yub}. The fact that, upon turning on $q$, $\tilde{q}$, the solutions become nonconstant was also explained there (and confirmed by numerical results for $N=3, k=1, r=2$). 
  The constant-$F$ solutions in this case are thus a set of measure zero on the moduli space, whose global structure 
 is unknown.  
  
  The third column refers to $Q \ge 1$, where $Q$ can also take all integer values. The observation of \cite{Anber:2025yub} that for $q, \tilde{q} \ne0$ the solutions become nonconstant applies also to this case. Regarding the global structure, we refer to the only reference we know of---the mathematical work of Braam, Todorov and Maciocia, for $SU(2)$ $Q=1$ instantons on $\T^4$ with a single twist ($n_{34}=1$ in our language). While it should have a relation to our $D$-brane picture, it is currently not understood. The class of solutions with $Q\in \Z$ should match to the ADHM one in an appropriate infinite volume limit.
  
  We also stress that the only solutions known explicitly, for all values of their moduli, are the constant-$F$ ones of eqns.~(\ref{SUNbackgroundIntro}) with gcd$(k,r)=r$.
The rest of the $(N,k,r)$-space of solutions on the tuned-$\T^4$, the nonconstant solutions with gcd$(k,r) \ne r$ are only known to leading order in the nonlinearity \cite{Anber:2025yub}, i.e. are obtained as small self-dual fluctuations around the constant-$F$ ones (as outlined in Section \ref{sec:missing}).  These correspond to giving infinitesimal expectation values of the moduli $q$, $\tilde{q}$ consistent with the local Higgs branch description of the moduli space of eqn.~(\ref{fterm2intro}). Figuring out the more general solutions or the global structure of their moduli is an outstanding task.

\subsection{Outlook}
\label{sec:outlook}
 
While we  already stated our satisfaction with the results obtained so far,  there remain many interesting and challenging (at least to the author) directions to pursue. We thus proceed with a wishlist. 

While restricting to a  4d EFT  does not help understand the global structure of the moduli space, a glimmer of  hope is  that the ease with which the local structure of the moduli space was determined may eventually be extended to other properties. Extracting the instanton profile is one of them. This was first done for instantons on $\R^4$ in \cite{Douglas:1995bn,Douglas:1996uz} using, e.g.~a $D_0$ brane probing  a $D_4$-$D_8$ instanton background, while  later approaches are in \cite{Hashimoto:2005qh,Billo:2002hm,Tong:2014cha}.
Our hope is that a combination of different tools will be useful to elucidate the moduli-space structure and shed light on 
the  nature of the space-time dependent solutions with the ``missing'' moduli turned on. This should help clarify the relation between the finite and infinite volume instantons and between the different incarnations of fractionally-charged objects mentioned in Section \ref{sec:history}.

Finally, for  a $\T^4$ whose shape does not obey (\ref{tuneintro}), there is a tachyon in the spectrum of open strings connecting the intersecting branes. The final point of tachyon condensation should be a minimum action space-time dependent instanton. In QFT, approximate analytic solutions were constructed via the ``$\Delta$-expansion''   \cite{GarciaPerez:2000aiw,Gonzalez-Arroyo:2019wpu, Anber:2023sjn} and one naturally wonders whether string theory tools can be useful in this regard.

\section{Field theory: embedding fractional instantons on $\T^4$ in $U(N)$}
\label{qftsetup}

\subsection{The Chern characters of the twisted $SU(N)\times U(1)/\mathbb Z_N$ bundle}
\label{sec11}

 We consider YM theory on $\mathbb T^4$ with a $SU(N)\times U(1)/\mathbb Z_N$ bundle. We take $\Omega_{\mu}$ and $\omega_\mu$ to be the $SU(N)$ and $U(1)$ transition functions, respectively, which satisfy the cocycle conditions
 \begin{eqnarray}\label{cocycle conditions for both}
 \nonumber
 \Omega_{\mu}(x+\hat e_\nu L_\nu)\Omega_\nu(x)&=&e^{i\frac{2\pi n_{\mu\nu}}{N}} \Omega_\nu(x+\hat e_\mu L_\mu)\Omega_\mu (x)\,,\\
  \omega_{\mu}(x+\hat e_\nu L_\nu)\omega_\nu(x)&=&e^{-i\frac{2\pi n_{\mu\nu}}{N}} \omega_\nu(x+\hat e_\mu L_\mu)\omega_\mu (x)\,,
 \end{eqnarray}
upon traversing $\mathbb T^4$ in any direction.  The vectors  $\{\hat e_\mu\}$ are unit normals in the $\mu=1,2,3,4$ directions, and $L_\mu$ are the length of cycles of $\mathbb T^4$. The integers $n_{\mu\nu}$ satisfy  $n_{\mu\nu}=-n_{\nu\mu}$ and are defined (mod $N$). Notice the negative sign difference in the $\mathbb Z_N$ phases of the two equations in (\ref{cocycle conditions for both}), which ensures  the combined transition functions satisfy proper $U(N)$ cocycle conditions. The $U(N)$ transition functions are
\begin{eqnarray}\label{UNtransition}
\Sigma_\mu (x) = \Omega_\mu(x) \omega_\mu(x),
\end{eqnarray} satisfying the cocycle condition
\begin{eqnarray}\label{UNcocycle}
 \Sigma_{\mu}(x+\hat e_\nu L_\nu) \Sigma_\nu(x)&=&  \Sigma_\nu(x+\hat e_\mu L_\mu) \Sigma_\mu (x)\,.
\end{eqnarray}
The $U(N)$ gauge field ${\cal A}$ obeys the boundary conditions
\begin{eqnarray}\label{BCS for A}
{\cal A}_\nu(x+ \hat e_\mu L_\mu)=\Sigma_\mu (x)({\cal A}_\nu(x) - i \partial_\nu) \Sigma_\mu^{-1}(x)~.
\end{eqnarray}
The $U(N)$ background  ${\cal A}_\mu$ can be split it into a $U(1)$ part, denoted by $a_\mu$, and an $SU(N)$  part $A_\mu$:
\begin{eqnarray}\label{UNbackground1}
{\cal A}_\mu = A_\mu + {\bf 1}_N \; a_\mu~,~ \text{tr} A_\mu = 0, ~ a_\mu = {1 \over N} \; \text{tr} {\cal A}.
\end{eqnarray}
Similarly, ${\cal F} = F + I_N d a \equiv F + I_N f$, with tr$F=0$ and $f =  {1 \over N} \; \text{tr} {\cal F}$ ($I_N$ is the unit matrix). The boundary conditions for $A$ and $a$ are  determined by $\Omega_\mu$ and $\omega_\mu$, respectively, as in (\ref{BCS for A}).

In  previous research (from the earliest \cite{tHooft:1981nnx,vanBaal:1984ar}, to \cite{GarciaPerez:2000aiw,Gonzalez-Arroyo:2019wpu} and the present-day  \cite{Anber:2022qsz,Anber:2023sjn,Anber:2024uwl,Anber:2024mco,Anber:2025yub,Anber:2025vjo}),  constant field strength $SU(N)$ instantons (and some properties of their non-constant deformations) on a twisted $\T^4$, characterized by topological charges $Q_{SU(N)}=r/N$ were extensively examined. As a quick reminder, one uses 't Hooft's idea \cite{tHooft:1981nnx} of embedding $SU(k)\times SU(\ell)\times U(1)$ within $SU(N)$, where $\ell+k=N$. 
To describe the solutions, we begin by defining the $U(1)$ generator
\begin{equation}\label{omega}
\omega=2\pi\mbox{diag}\left[\underbrace{\ell, \ell,...,\ell}_{k\, \mbox{times}},\underbrace{ -k,-k,...,-k}_{\ell\,\mbox{times}}\right],~ \ell + k = N, \end{equation} and further introduce the matrices $P_\ell$ and $Q_\ell$,  the $\ell\times \ell$ shift and clock matrices:
\begin{eqnarray}\label{pandq}
P_\ell=\gamma_\ell\left[\begin{array}{cccc} 0& 1&0&...\\ 0&0&1&...\\... \\ ...&  &0&1 \\1&0&...&0\end{array}\right]\,,\quad Q_\ell=\gamma_\ell\; \mbox{diag}\left[1, e^{\frac{i 2\pi}{\ell}}, e^{2\frac{i 2\pi}{\ell}},...\right]\,,
\end{eqnarray}
which satisfy the relation $P_\ell Q_\ell=e^{i\frac{2\pi}{\ell}}Q_\ell P_\ell$. The factor $\gamma_\ell \equiv e^{\frac{i\pi (1-\ell)}{\ell}}$ ensures that $\det Q_\ell=1$ and $\det P_\ell=1$.
$P_k, Q_k$ and $\gamma_k$ are defined similarly. 

Then,
recalling the cocycle conditions (\ref{cocycle conditions for both}), we take the nontrivial twists $n_{\mu\nu}$ to be\footnote{We stress  that all transition functions given below have nontrivial dependence on the absolute value of $r$, not only modulo $N$.}
\begin{eqnarray}\label{twists}
n_{12}=-r\,,\quad n_{34}=1\,, \;\; r \in \mathbb{N}.
\end{eqnarray}
The $SU(N)$ transition functions are given in a gauge where the field strength itself  is manifestly constant (rather than only the gauge invariants associated with it)
\cite{tHooft:1981nnx}%
 \begin{eqnarray}
\nonumber
\Omega_1&=&\left[\begin{array}{cc}P_k^{-r}e^{i2\pi \ell r \frac{x_2}{Nk L_2}}&0\\0& e^{-i 2\pi r\frac{x_2}{NL_2}}I_\ell\end{array}\right],~
\nonumber
\Omega_2=  \left[\begin{array}{cc}Q_k&0\\0& I_\ell\end{array}\right],\\
\Omega_3&=&\left[\begin{array}{cc} e^{i2\pi  \frac{x_4}{N L_4}} I_k&0\\0& e^{-i 2\pi k\frac{x_4}{N \ell L_4}}P_\ell\end{array}\right],~~~~  ~
\Omega_4=  \left[\begin{array}{cc}I_k&0\\0& Q_\ell\end{array}\right].
\label{the set of transition functions for Q equal r over N, general solution}
\end{eqnarray}
These are combined  \cite{Anber:2024uwl} with the following abelian transition functions %
\begin{eqnarray}\label{u1transition}
\omega_1=e^{i\frac{2\pi(r-q_1 N)x_2}{N L_2}}\,,\quad \omega_3=e^{-i\frac{2\pi(1+q_3 N)x_4}{N L_4}}\,, \quad \omega_2=\omega_4=1\,,
\end{eqnarray}
where, for definiteness, we take $q_1$ and $q_3$ to be nonnegative integers.\footnote{In our moduli space analysis,  beginning with Section \ref{sec:numberofintersect}, we will take $q_1=q_3=0$.} It is straightforward to check that the $SU(N)$ and $U(1)$ transition functions obey (\ref{cocycle conditions for both}) with $n_{\mu\nu}$ from (\ref{twists}).  

The $U(1)$ transition functions (\ref{u1transition}) determine the fluxes of $f = d a$ through the $12$ and $34$ planes (equal to $1/N$-th of the corresponding first Chern characters $\text{ch}_1({\cal F})$):
\begin{eqnarray}\label{fluxU1}
 \int\limits_{\T^2_{(x_1, x_2)}} {\text{ch}_1({\cal{F}})  \over N}&=&  \int\limits_{\T^2} dx_1 dx_2 {f_{12} \over 2 \pi} =  -  {r  \over N } + q_1, \nonumber \\
  \int\limits_{\T^2_{(x_3, x_4)}} {\text{ch}_1({\cal{F}})  \over N} &=& 
 \int\limits_{\T^2} dx_3 dx_4 {f_{34} \over 2 \pi}  =  {1\over N}+ q_3, \end{eqnarray}
as well as the $U(1)$ topological charge: \begin{eqnarray} \label{QU1}
\quad Q_{U(1)}&=&{1 \over 8 \pi^2}  \int_{\T^4} f \wedge f  = -\left(\frac{r}{N}-q_1\right)\left(\frac{1}{N}+q_3\right)~.\end{eqnarray}
The $U(N)$ topological charge, or second Chern character $\text{ch}_2({\cal F})$, is
  \begin{eqnarray}\label{secondchern}
 \int\limits_{\T^4} \text{ch}_2({\cal F}) &=& {1 \over 8 \pi^2} \int\limits_{\T_4} \text{tr} {\cal F} \wedge {\cal F} = {1 \over 8 \pi^2} \int\limits_{\T_4} \text{tr} { F} \wedge {  F}  + {N \over 8 \pi^2} \int\limits_{\T^4} f \wedge f \nonumber \\
  &=& Q_{SU(N)} + N Q_{U(1)} =N q_1 q_3 + q_1 - r q_3,\end{eqnarray}
and is expressed in terms of $Q_{U(1)}$ and $Q_{SU(N)}$. To obtain the above result, we used (\ref{QU1}) and, to calculate $Q_{SU(N)}$, we used the   $SU(N)$ transition functions (\ref{the set of transition functions for Q equal r over N, general solution}), with the result 
\begin{eqnarray}\label{qsun}
Q_{SU(N)}&=&\frac{r}{N}.
\end{eqnarray}
To find (\ref{qsun}), one uses the fact that the integrand in $Q_{SU(N)}$ is a total derivative and repeatedly integrates by parts, using the transition functions at each step, to obtain an expression in terms of $\Omega_\mu$ only,  whose evaluation gives the above result.\footnote{For the reader who wants to repeat the calculation, we note that the  $x_\mu$-independence of  $\Omega_{2,4}$ helps  speed it up, reducing the answer to $Q_{SU(N)} = - {1 \over 4 \pi^2} \int_{\T^2_{(x_2, x_4)}} \; \text{tr}\; \Omega_1^{-1} d \Omega_1  \Omega_3^{-1} d \Omega_3$, easily seen to equal (\ref{qsun}). An alternative is to find a simple background consistent with the boundary conditions (\ref{BCS for A}), e.g.~the one of eqn.~(\ref{SUNbackground12}), and calculate its topological charge.}
We stress that for $q_1=q_3=0$, the $SU(N)$ and $U(1)$ (times $N$) topological charge are equal and opposite. Generally, a nonzero fractional $SU(N)$ topological charge is imposed by the nonzero fractional $U(1)$ fluxes (\ref{fluxU1}). 

Finally, to avoid any confusion, we stress that the value of $r$ can be any positive integer, even such that (\ref{qsun}) is a natural number, including $Q_{SU(N)}=1$. The reader may recall the well known fact that with periodic boundary conditions, i.e. transition functions with all $n_{\mu\nu}=0\; (\text{mod} \; N)$, there are no charge-1 instantons in $SU(N)$ on $\T^4$ (this follows from the Nahm transform \cite{Braam:1988qk}). That there is no contradiction follows from observing that even with $r$ proportional to $N$ and thus a trivial $n_{12}$, the $n_{34}$ twist in  (\ref{twists}) is still nontrivial. Thus,  arbitrary integer-charge solutions exist in our setup.

\subsection{Constant flux backgrounds with $Q_{SU(N)}$$=$${r/N}$, $r$$\in$${\mathbb{N}}$, and the  BPS conditions}
\label{sec:constant}

We already mentioned the class of explicit backgrounds satisfying the boundary conditions with transition functions given above---the ones of constant fluxes on $\T^4$ \cite{tHooft:1981nnx}. As discussed below, these backgrounds are self-dual for appropriately tuned sides of $\T^4$.

We first give the field strength for the $U(N)$ ($\cal{F}$) and $SU(N)$ ($F$) gauge fields (\ref{UNbackground1}) in a compact matrix form and later give the vector potential in a convenient index notation.
The  $U(N)$ field strengths (whose vector potentials are in~(\ref{UNbackground2}, \ref{UNbackground3}) below)  are constant, with only nonzero components on the diagonal:
\begin{eqnarray}\label{UNbackground12}
{\cal F}_{12} &=&\left(\begin{array}{cc}{2 \pi (k q_1 - r) \over k L_1 L_2} {I}_k & 0 \cr 0 & {2 \pi q_1 \over L_1 L_2} {I}_\ell \end{array}\right), ~~
{\cal F}_{34} =\left(\begin{array}{cc}{2 \pi q_3 \over L_3 L_4} {I}_k & 0 \cr 0 & {2 \pi (\ell q_3 +1) \over \ell L_3 L_4} {I}_\ell \end{array}\right). 
\end{eqnarray}
Imposing the $U(N)$ self-duality (BPS) condition demands that the torus sides are tuned
\begin{eqnarray}\label{UNBPS}
U(N)\; \text{BPS}:~~{L_1 L_2 \over L_3 L_4} = {   k q_1 - r \over k q_3} = {\ell q_1 \over \ell q_3 + 1} \implies k q_1 = r \ell q_3+r. \end{eqnarray}
As indicated above, the $U(N$) BPS condition imposes constraints on the integer $U(1)$-fluxes $q_{1,3}$. Note in particular, that for specific choices of $r, k$ there are $U(N)$ self-dual backgrounds with only $q_1$ or $q_3$ nonzero, but not both.

We can also project out the $U(1)$ to find the $SU(N)$ field strength from (\ref{UNbackground12}), obtaining for $F = {\cal F} - {I_N \over N}  \tr {\cal F}$,
\begin{eqnarray}\label{SUNbackground12}
{ F}_{12} &=&\left(\begin{array}{cc} - {2 \pi \ell r   \over N k L_1 L_2} {I}_k & 0 \cr 0 & {2 \pi r  \over N L_1 L_2} {I}_\ell \end{array}\right), ~~{ F}_{34} = \left(\begin{array}{cc} - {2 \pi   \over N L_3 L_4} {I}_k & 0 \cr 0 & {2 \pi k \over N \ell L_3 L_4} {I}_\ell \end{array}\right)~, 
\end{eqnarray}
implying the $SU(N)$ BPS condition:
\begin{eqnarray}\label{SUNBPS}
SU(N)\; \text{BPS}:~~{L_1 L_2 \over L_3 L_4} = {  \ell r \over k },
\end{eqnarray}
clearly, a  less restrictive condition than the $U(N)$ self-duality (\ref{UNBPS}) and independent of $q_1, q_3$.

We now continue with the  
 $U(N)$  vector potentials
 that give rise to (\ref{UNbackground12}, \ref{SUNbackground12}). These 
  obey the boundary conditions with the transition functions (\ref{the set of transition functions for Q equal r over N, general solution}, \ref{u1transition}).  For now, we do not include moduli. Recalling the expression for $\omega$ from (\ref{omega}), we write the vector potentials as%
\begin{eqnarray}\label{the full abelian bck general r}
\nonumber
A_2&=&-\omega\left(\frac{r x_1}{Nk L_1 L_2}\right)\,,\quad  A_4=-\omega\left(\frac{x_3}{N\ell L_3 L_4}\right)\,, \quad A_1=A_3=0\,,\\
a_2&=&- I_N \frac{2\pi (r-q_1 N)x_1}{N L_1L_2}\,,\quad a_4= I_N \frac{2\pi (1+q_3 N)x_3}{N L_3L_4}\,, \quad a_1=a_3=0\,.
\end{eqnarray} 

To continue, we now switch to an index notation, to be used throughout the paper. We split the $SU(N)$ indices into $C', D',... = 0,..., k-1$ and $C,D,... = 0,... \ell-1$. This notation  will allow us to include the general allowed moduli. Thus,  we rewrite the $SU(N)$ part of the background (\ref{the full abelian bck general r}) as follows. The $k\times k$ components of the background are:
\begin{eqnarray}
\label{SUkbackground}
A_{1 \; C'D'} &=& - \delta_{C'D'} \; 2 \pi \ell \; \phi_{1 \; C'},~
A_{2 \; C'D'} = -  \delta_{C'D'} \; 2 \pi \ell\;( {r x_1 \over N k L_1 L_2}  + \phi_{2 \; C'})\\
A_{3 \; C'D'} &=&-  \delta_{C'D'} \; 2 \pi \ell \;\phi_{3 \; C'},~
A_{4 \; C'D'} = - \delta_{C'D'} \; 2 \pi \ell \;({x_3 \over N \ell L_3 L_4}+ \phi_{4 \; C'}) \nonumber
\end{eqnarray}
where the allowed $SU(k) \times U(1) \subset SU(N)$ moduli  are labelled $\phi_{\mu\; C'}$. Their properties are explained in the next paragraph, see (\ref{SUNmoduli}).
The remaining part of the $SU(N)$ background is in the $ \ell \times \ell$ part of   $SU(N)$: 
\begin{eqnarray}
\label{SUlbackground}
A_{1 \; CD} &=&   \delta_{CD} \; 2 \pi k \;  \tilde\phi_{1},~
A_{2 \; CD} = \delta_{CD} \; 2 \pi k \;( {r x_1 \over N k L_1 L_2}  + 
\tilde\phi_{2})\\
A_{3 \; CD} &=&  \delta_{CD} \;  2 \pi k\;  \tilde\phi_{3},~
A_{4 \; CD} =   \delta_{CD}\; 2 \pi k \; ({x_3 \over N \ell L_3 L_4}+ \tilde\phi_{4}),\nonumber
\end{eqnarray}
where $\tilde\phi_\mu$ is defined in (\ref{SUNmoduli}) below (noting that 
it ensures tracelessness, $A_{1\; CC} + A_{1\; C'C'}=0$, etc.)
 The moduli $\phi_{\mu \; C'}$, $C' = 1,...,k$, are not all independent, but, in order to be consistent with the transition functions,  are subject to the identifications
\begin{eqnarray} \label{SUNmoduli}
\phi_{\mu \; C'} &=& \phi_{\mu \; C'-r (\text{mod} \; k)} \equiv \phi_{\mu\; [C'-r]_k}~,~\text{and we define, for use below:}\nonumber \\
\tilde\phi_{\mu} &\equiv& {1 \over k} \sum\limits_{C'=1}^k \phi_{\mu \; C'},
\end{eqnarray}
where the latter condition ensures tracelessness of $SU(N)$. 
The identification $[C'-r]_k = C'$ implies that are gcd$(k,r)$ moduli $\phi$ for each $\mu$, for a total of $4$gcd$(k,r)$ $SU(N)$ moduli.

 The $\phi_{\mu \; C'}$ from (\ref{SUNmoduli}) are the most general  constant connections  in $SU(N)$ that can be added to the background (\ref{SUkbackground}, \ref{SUlbackground}). These do not raise the action above the BPS limit (which is saturated by the constant-$F$ backgrounds when the torus sides are appropriately tuned, see (\ref{SUNBPS})). This was shown in \cite{Anber:2023sjn}, but because of the importance of the moduli in this paper, we quickly repeat the argument here. A constant contribution to (\ref{SUkbackground}, \ref{SUlbackground}) has to commute with all transition functions (\ref{the set of transition functions for Q equal r over N, general solution}). Starting with $\Omega_{2}$ and $\Omega_4$, we note that a matrix that commutes with them has to be diagonal. Further, commutativity with $\Omega_3$ demands that the last $\ell$ eigenvalues be all the same, while commutativity with  $\Omega_1$ permits the first $k$ eigenvalues to be only identical in groups of gcd$(k,r)$ ones. This yields the result given in (\ref{SUNmoduli}), with the notation $[C'-r]_k \equiv C'- r (\text{mod} k)$. Adding the $U(1)$ moduli $z_\mu$ from (\ref{U11background}) gives  thus a total of $4$ gcd$(k,r)$ $+4$ moduli.  

The abelian background responsible for the fluxes (\ref{fluxU1}), including the $U(1)$ moduli $z_\mu$ is
\begin{eqnarray}\label{U11background}
 a_1&=& I_N  \;2 \pi z_1,~
 a_2=-I_N \frac{2\pi (r-q_1 N)x^1}{N L_1L_2} + I_N \;2 \pi z_2\\
 a_3&=& I_N\; 2 \pi z_3, ~
 a_4=I_N \frac{2\pi x^3 (1+q_3 N)}{N L_3L_4} + I_N\; 2 \pi z_4. \nonumber
\end{eqnarray}

{\flushleft\bf{Summary of the background, including moduli, in index notation:}} Here, we combine (\ref{U11background})  with the $SU(N)$ backgrounds, to give the $U(N)$ background ${\cal A}$. 
Its $k\times k$  components are, with the $SU(N)$ moduli $\phi_\mu$ obeying (\ref{SUNmoduli}):
\begin{eqnarray}\label{UNbackground2}
{\cal A}_{1 \; C'D'} &=& \delta_{C'D'} 2 \pi(z_1 - \ell \phi_{1 \; C'}),~
{\cal A}_{2 \; C'D'} = \delta_{C'D'} 2 \pi(z_2 - \ell \phi_{2 \; C'} + {k q_1 - r \over k L_1 L_2} \; x_1), \\
{\cal A}_{3 \; C'D'} &=& \delta_{C'D'} 2 \pi(z_3 - \ell \phi_{3 \; C'}),~{\cal A}_{4  \; C'D'} = \delta_{C'D'} 2 \pi(z_4 - \ell \phi_{4 \; C'} + {q_3  \over L_3 L_4} \; x_3)~, \nonumber
\end{eqnarray}
while the $\ell \times \ell$ components are:
\begin{eqnarray}
\label{UNbackground3}
{\cal A}_{1 \; CD} &=& \delta_{CD'} 2 \pi(z_1 + k  \tilde\phi_{1}),~
{\cal A}_{2 \; CD} = \delta_{CD} 2 \pi(z_2 + k \tilde\phi_{2} + { q_1 \over   L_1 L_2} \; x_1), \\
{\cal A}_{3 \; CD} &=& \delta_{CD} 2 \pi(z_3 + k\tilde\phi_{3 }),~
{\cal A}_{4  \; CD} = \delta_{CD} 2 \pi(z_4 + k \tilde\phi_{4} + {\ell q_3 + 1  \over \ell L_3 L_4} \; x_3)~.\nonumber
\end{eqnarray}
 The backgrounds in the form given above are of most utility for finding $D$-brane picture.
 
\subsection{The ``missing moduli'' puzzle: a review of its resolution in  field theory}
\label{sec:missing}

The $4$gcd$(k,r)$ moduli appearing in the $SU(N)$ background of eqns.~(\ref{SUkbackground}, \ref{SUlbackground}), the constant holonomies $\phi_{\mu C'}$ obeying (\ref{SUNmoduli}), do not saturate the index theorem result, $4r$  \cite{Schwarz:1977az,Weinberg:1979ma,Taubes:1982qem} unless gcd$(k,r)=r$. To investigate this,  in \cite{Anber:2025yub}  we  proceeded  to study the self-dual fluctuations around the solutions (\ref{SUkbackground}, \ref{SUlbackground}). The description given below  only aims to present the idea and result. 

We denote the  background (\ref{SUkbackground}, \ref{SUlbackground}) by $A$ and consider  general $SU(N)$ fluctuations $a$, $A'=A+a$, around it. The fluctuations $a$ obey  the background gauge condition $D(A)_\mu a_\mu = 0$, with $D(A)$  the background covariant derivative. They also obey boundary conditions on $\T^4$ consistent with the transition functions. 
We next demand that $A+a$ be self dual, i.e.
$F(A+a) = *F(A+a)$, where of course $A$ obeys $F(A) = *F(A)$. The solutions of this equation explore the neighborhood  (if $a$ is small) of $A$ in the space of self-dual fields with the same minimal action. This self-duality equation  
is quadratic in  $a$ and can be studied using an expansion in the nonlinearity \cite{Schwarz:1977az,Weinberg:1979ma,Taubes:1982qem}.

Proceeding thus, to linear order in $a$, the solution of the self-duality equation, in addition to constant functions that give rise to the moduli $\phi_{\mu C'}$, was found to contain $2r$  functions (space-time dependent) on the torus, explicitly given in refs.~\cite{Anber:2023sjn,Anber:2025yub}. A general small fluctuation is then described by a linear combination of these functions, with complex coefficients, denoted by ${\cal{C}}_{2}^{A}$ and ${\cal{C}}_{4}^{A}$, $A=1,...r$; we shall see that these coefficients contain   the ``missing moduli.''   We note that $2r$ is precisely the number of complex fields in the bifundamental hypers $q_i^a, \tilde{q}_a^i$ ($a=1,...,r/\text{gcd}(k,r)$, $i=1,...,\text{gcd}(k,r)$)  appearing at the brane  intersections and in eqn.~(\ref{fterm2intro}).  To further strengthen the relation between the two,  the set of  indices  $A=1,...,r$ is split into gcd$(k,r)$ groups of $r/\text{gcd}(k,r)$ integers, as shown in \cite{Anber:2023sjn}. These groups of integers 
are labeled by $S_j$, $j=1,...,\text{gcd}(k,r)$. 
 A concise summary of their  properties\footnote{For the curious, the precise definition of $S_j$ is, using, once again, the shorthand notation $[a]_p = a \;(\text{mod}\;p)$, with $[p]_p\equiv p$:
 \begin{equation}\nonumber
S_j = \bigg\{ [[j+nr]_k+ pk]_r, \text{for} \; n = 1,...\frac{k}{{\rm gcd}(k,r)},  \text{and} \; p = 1,...,{r\over {\rm gcd}(k,r)} \bigg\}, \; j=1,...,\text{gcd}(k,r), \end{equation} where repeated entries obtained from the   above  are identified;  each set $S_j$ has $r \over {\rm gcd}(k,r)$ elements and 
 the union of all sets $S_j$ is the set $ \{1,...,r\}$ as shown in (\ref{sets}). To verify  this, one needs to patiently examine the definition of $S_j$ above;  for a derivation see the voluminous Appendices A, B of \cite{Anber:2023sjn}.}
 is all we will need here  \begin{eqnarray}\label{sets}
  ~|S_j| = {r \over \text{gcd}(k,r)},\; ~ S_i \cap S_{j \ne i} = \varnothing~, \; S_1 \cup S_2 \cup ... \cup S_{\text{gcd}(k,r)} = \{1,..., r\}.  
\end{eqnarray}
The final step    is to require consistency of the perturbative expansion of the self-duality equation $F(A+a) = *F(A+a)$,  too long to explain here.\footnote{See Sections 3.1 and 4.1 in \cite{Anber:2025yub}.} The end result is that the coefficients of the perturbations obey a set of constraints:
\begin{eqnarray}\label{constraintssubbed}
\sum\limits_{A \in S_j} {\cal C}_2^{A} \;{\cal C}_2^{* \; A} - {\cal C}_4^{A} \;{\cal C}_4^{* \; A} &=&0, ~\forall \; j=1,...,\text{gcd}(k,r), ~\text{$g$ real constraints},
 \\
\sum\limits_{A \in S_j} {\cal C}_2^{A}\; {\cal C}_4^{* \; A} &=&0~,  ~\forall \; j=1,...,\text{gcd}(k,r), ~ ~\text{$2g$ real constraints}.\label{bottom} \end{eqnarray}
Apart from the complicated splitting of the set of integers $\{1,...,r\}$ into  the sets of indices defined in (\ref{sets}), these are precisely the $D$-  and $F$-term conditions, given in eqns.~(\ref{fterm2intro}) of the Introduction, or eqns.~(\ref{fterm2}, \ref{dterm2}) of Section \ref{sec:u1branch}. Explicitly, the $D$- and $F$-terms match the above eqns.~(\ref{constraintssubbed}, \ref{bottom}) upon identifying  the sets $C_2^A\vert_{A \in S_j} \leftrightarrow q_j^a\vert_{a=1,...,r/\text{gcd}(k,r)}$ and $C_4^{* A}\vert_{A \in S_j} \leftrightarrow \tilde{q}_a^j\vert_{a=1,...,r/\text{gcd}(k,r)}$. 
In Section \ref{sec:u1branch} below, eqns.~(\ref{constraintssubbed}, \ref{bottom}) will be seen to arise in a much more straightforward way from the $D$-brane construction.

{\flushleft{Before}} we continue with the $D$-brane setup,   two more remarks on the QFT results are due:
\begin{enumerate}

\item The QFT study of the moduli space linearized around (\ref{SUkbackground}, \ref{SUlbackground}) reveals that when the moduli ${\cal{C}}_{2}^A, {\cal{C}}_{4}^A$ are turned on, the instanton background becomes space-time dependent (in the same manner as eqns.~(\ref{fterm2intro}),  eqns. (\ref{constraintssubbed}) only allow nonzero values for the moduli for $r \ne \text{gcd}(k,r)$). Thus, the constant-$F$ solutions of 't Hooft are a measure zero set in the moduli space, for gcd$(k,r) \ne r$, the case which includes all solutions with $Q\ge 1$. In the language of the brane construction ``Higgs branch'' equations, 
't Hooft's constant-$F$ solutions correspond to taking $q = \tilde{q}=0$ in the form (\ref{fterm2intro}), or, equivalently, $\chi^a_j=0$ for the form (\ref{hyperkintro}).

 The $x_\mu$-dependence of $Q={r/N}$ solutions with gcd$(k,r)\ne r$ was,  also in \cite{Anber:2025yub}, subjected to a numerical lattice test, to leading order in the nonlinearity, for the simplest case with gcd$(k,r)\ne r$: $N=3$, $k=1$, $r=2$.
  Higher orders in the perturbation theory have not been pursued in QFT, and are challenging, for reasons explained in \cite{Anber:2025yub}. In view of the easy derivation of (\ref{constraintssubbed}) from the $D$-brane picture, it would be of interest to pursue this further.
\item As already noted, eqns.~(\ref{constraintssubbed}, \ref{bottom}) fix ${\cal{C}}^A_{2,4}=0$ when gcd$(k,r)=r$. We showed in \cite{Anber:2025yub} that this remains so to all orders in the nonlinearity expansion. Thus, the $4 r$ holonomies $\phi_{\mu C'}$ comprise all the moduli (notice that this is only relevant for some  of the $r<N$ backgrounds). 

For  $Q_{SU(N)}<1$ solutions with gcd$(k,r)=r$, for $k=r$, the moduli space structure was determined globally in \cite{Anber:2024mco} (assuming the absence of disconnected components). Briefly, the idea was to consider (arbitrary powers of) the fundamental representations winding Wilson loops in the background of the constant-$F$ solution and demand that the range of $\phi_{\mu C'}$ be such that these Wilson loops vanish when integrated over the moduli space (a condition which follows from the preservation of center symmetry, see Section 3 there).
The resulting  global structure of the moduli space was determined to be 
\begin{eqnarray}\label{modspacegcd=r}
\Gamma = \left(\prod\limits_{\mu=1}^4 {(\S^1)^r \over \Z_r}\right)/S_r.
\end{eqnarray}
The action of $\Z_r$ is free (without fixed points) and $S_r$ is a permutation acting simultaneosly on all four elements of the product. We note that the space defined in  (\ref{modspacegcd=r}) is free of singularities.
The metric and volume form are given in \cite{Anber:2024mco}. This was the moduli space we integrated over in order to compute the $r$-point ($r<N$) gaugino condensate $\langle(\tr \lambda\lambda)^r\rangle$,  obtaining exactly the result of \cite{Dorey:2002ik} which used the ADHM construction.\footnote{The only moduli for gcd$(k,r)=r$ appear in the dual brane picture as $\tilde{\T}^4$ brane positions and worldvolume Wilson lines, see Figure \ref{brane5}. These are completely equivalent to the $\phi_{\mu C'}$ moduli (\ref{SUNmoduli}) from QFT, see (\ref{A1primeK}, \ref{A1primeL}) and Section \ref{sec:branesdual} for explicit expressions. The global determination of their moduli space  structure is thus equivalent to the one in \cite{Anber:2024mco}. } The agreement of the $\T^4$ calculation of the order-$r$ gaugino condensate of that reference   with the $\R^4$ determination of \cite{Dorey:2002ik}  gives strong evidence that the moduli space structure we found is correct.
\end{enumerate}
\section{$\mathbf{D}$-branes: instanton embedding, $T$-duality, and the moduli space}

The embedding of 't Hooft's constant-$F$ fractional instantons in string theory also proceeds via a $U(1)$ twist compensating the $SU(N)$ one, as we have considered in this paper. Various aspects of similar, but not identical, constructions were studied  in refs.~\cite{Berkooz:1996km,Taylor:1996ik,Guralnik:1997sy,Hashimoto:1997gm,Taylor:1997dy}, among others, but  the moduli space of fractional instantons was not of concern at the time and was not studied. The discussion below  also serves  to illustrate the difference between the string embedding of $k$ integer-charge BPST-instantons (leading to a brane realization of the ADHM construction \cite{Witten:1994tz,Douglas:1996uz}, see \cite{Tong:2005un} for an Introduction) and the embedding of the constant flux fractional instantons. 

Consider the rectangular four torus $\tilde{\T}^4$  with coordinates $y_{1,...,4}$ of periods $\hat{L}_{1,...4}$. Consider also the four torus $\T^4$, the one of Section \ref{qftsetup}, of coordinates $x_{1,...,4}$ of periods $L_{1,...,4}$. For use below, we take them to be related by $T$-duality in the $2$ and $4$ directions, for which we have the standard relations between the sides of the $\T^4$ and $\tilde\T^4$, with $\alpha'$ being the string-length squared:
\begin{eqnarray}
\tilde{\T}^4_{y_\mu \in [0, \tilde{L}_\mu]} &\leftrightarrow& \T^4_{x_\mu \in [0, {L}_\mu]}:~
\hat L_1 = L_1,
\hat L_2 = {4 \pi^2 \alpha' \over L_2}~,    \hat L_3 =  L_3~,
\hat L_4 =  {4 \pi^2 \alpha' \over L_3}~. \label{tdual1}
\end{eqnarray}

Our strategy is to start with  $D_{p+4}$-branes\footnote{As explained in footnote \ref{footnoteatwill}, a value for $p$ can be chosen for convenience.} of type-II string theory, with four of its worldvolume directions wrapped on $\T^4$, with the   background (\ref{UNbackground2}, \ref{UNbackground3}) turned on. We then perform a $T$-duality transformation to find an equivalent picture in terms  of intersecting stacks of wrapped $D_{p+2}$-branes  \cite{Taylor:1996ik,Taylor:1997dy}. Our goal is to study how the unusual properties of the moduli space of fractional instantons on the tuned $\mathbb{T}^4$ (discussed in  \cite{Anber:2025yub} and reviewed in Section \ref{sec:missing}) appear in the $D$-brane setup. We shall see that some properties of the moduli space---found after rather long and laborious calculations in QFT---appear quite simply in string theory.

\subsection{A convenient gauge for $T$-duality}
\label{sec:convenient}

We start with the 4d gauge bundle on $\T^4$ described in detail in Section \ref{qftsetup}. The $\T^4$  coordinates are $x_{1},...,x_{4}$ and its periods--$L_{1},...,L_{4}$.
We shall interpret this background  as a field configuration on the worldvolume of $N$ $D_{p+4}$ branes wrapped on $\T^4$. We then perform $T$-duality to a brane-at-angle configuration on $\tilde \T^4$ and use it to study the moduli space.
The bundle of interest, of Section \ref{qftsetup}, has fractional $U(1)$ fluxes through the $12$ and $34$ planes, which, as described there, enforce fractional topological $SU(N)$ charge.
As before, we denote the $U(N)$ transition functions by $\Sigma_\mu = \Omega_\mu \omega_\mu$, with $\Omega_\mu$ given in (\ref{the set of transition functions for Q equal r over N, general solution}) and $\omega_\mu$ in (\ref{u1transition}): 
 \begin{eqnarray}
\nonumber
\Sigma_1(x_2)&=& e^{ - i\frac{2\pi  q_1 x_2}{  L_2} }\; \left[\begin{array}{cc}P_k^{-r}e^{i2\pi   r \frac{x_2}{ k L_2}}&0\\0& I_\ell\end{array}\right], ~
\Sigma_2 =  \left[\begin{array}{cc}Q_k&0\\0& I_\ell\end{array}\right],\\
\Sigma_3(x_4)&=&  e^{-i\frac{2\pi q_3  x_4}{  L_4}} \left[\begin{array}{cc}  I_k&0\\0& e^{-i 2\pi  \frac{x_4}{  \ell L_4}}P_\ell\end{array}\right],~~~~ \Sigma_4    = \left[\begin{array}{cc}I_k&0\\0& Q_\ell\end{array}\right].
\label{UNtransition1}
\end{eqnarray}
These obey the proper $U(N)$ cocycle conditions (\ref{UNcocycle}); $P_\ell$ and $Q_\ell$ are the shift- and clock-matrices from (\ref{pandq}),  likewise defined for $\ell \rightarrow k$.
The  $U(N)$  gauge backgrounds consistent with the transition functions (\ref{UNtransition1}) are given in (\ref{UNbackground2}, \ref{UNbackground3}). The formulae there include all $SU(N)$  moduli  $\phi_{\mu C'}$, obeying  (\ref{SUNmoduli}), and the $U(1)$ moduli $z_\mu$  from (\ref{U11background}). 

We shall perform $T$-duality in the $x_2$ and $x_4$ directions, with  the background fields (\ref{UNbackground2}, \ref{UNbackground3}) turned on in the $D_{p+4}$-brane $U(N)$ worldvolume theory. It is well known \cite{Taylor:1996ik, Taylor:1997dy} that $T$-duality is most easily performed in directions where the transition functions are unity. 
Thus, we first  perform a gauge transformation $g(x)$ on the background and the transition functions. We recall that  consistency with the boundary conditions requires that these transform as
$\Sigma_\mu'(x) =  g(x + L_\mu) \Sigma_\mu(x) g(x)^{-1}$, ${\cal A}_\mu'(x) = g(x) ({\cal A}_\mu(x) - i d) g(x)^{-1}$. 
To diagonalize $\Sigma_2$ and $\Sigma_4$, we take $g(x)$ 
to be the $SU(N) $matrix: \begin{equation}\label{gauge12}
g(x) =  \left[\begin{array}{cc}Q_k^{- {x_2 \over L_2}}&0\\0& Q_\ell^{-{x_4 \over L_4}}\end{array}\right].
\end{equation}
Recall from (\ref{pandq}) that $Q_{k, \ell}$ are  unimodular diagonal matrices. The fractional powers in (\ref{gauge12}) are defined as $Q_k^{\alpha} =   \gamma_k^\alpha \; \text{diag}(1, e^{i {2 \pi \over k} \alpha}, e^{i {2 \pi \over k}2 \alpha}, \ldots, e^{i {2 \pi \over k}(k-1) \alpha})$, and similarly for $k \rightarrow \ell$.  The new transition functions $\Sigma_2'$ and $\Sigma_4'$ are then obviously trivial:  \begin{eqnarray}\label{newgauge24}
\Sigma_2' =  \left[\begin{array}{cc}I_k&0\\0& I_\ell\end{array}\right],~ \Sigma_4' =\left[\begin{array}{cc}I_k&0\\0& I_\ell\end{array}\right]. \end{eqnarray}
To
 find  $\Sigma_1'$ and $\Sigma_3'$, we introduce index notation for the $C'$-th eigenvalue of $Q_k$:
$
( Q_k^\alpha)_{B'C'} = \delta_{B'C'} \gamma_k^\alpha\; e^{i {2 \pi \over k} (C'-1) \alpha}~,~C'=1,...,k, 
 $
with $(Q^\alpha_\ell)_{BC}$ is given by the same expression with $k \rightarrow \ell, B',C' \rightarrow B,C$. We also write $P_k$ in matrix form as
$(P_k^n)_{A'B'} = \gamma_k^n \delta_{A', [B'-n]_k} = \gamma_k^n \delta_{[A'+n]_k, B'}$, where 
$[A^\prime]_k \equiv A' (\text{mod} \; k) \in \{1, 2,...,k\}$, i.e.~$[k]_k=[0]_k =k$.
We obtain,
$
(Q_k^{- \alpha} P_k^n Q_k^\alpha)_{B'C'} = e^{- i {2 \pi \over k}(B'-C') \alpha} (P^n_k)_{B'C'} = \gamma_k^n e^{- i {2 \pi \over k}(B'-C') \alpha} \delta_{B' [C'-n]_k}~.
$
Armed with this, we can write $\Sigma_1'$ and $\Sigma_3'$, whose $k \times \ell $ and $\ell \times k$ blocks vanish (as in (\ref{UNtransition1})), and the rest is most conveniently written in index notation: \begin{eqnarray}\label{newgauge13}
( \Sigma_1' )_{B'C'} &=&  \delta_{B' [C'+r]_k} \gamma_k^{-r} e^{- i 2 \pi {x_2 \over L_2}({[C'+r]_k - C' - r \over k} + q_1)}, ~~
( \Sigma_1' )_{BC} =  \delta_{BC} e^{- i {2 \pi} q_1 {x_2 \over L_2}}, \\
(\Sigma_3')_{B'C'} &=& \delta_{B'C'} e^{- i {2 \pi} q_3 {x_4 \over L_4}}, ~~ \qquad \qquad \qquad \qquad (\Sigma_3')_{BC}    = \delta_{B [C-1]_\ell} \gamma_\ell \;e^{- i 2 \pi {x_4 \over L_2}({[C-1]_\ell - C + 1\over \ell} + q_3)}. \nonumber \end{eqnarray}
The $k \times k$ block of $\Sigma_1'$ is similar to a $P_k^{-r}$ permutation matrix, as in (\ref{UNtransition1}), but with an extra $x_2$-dependent phase for  $C'$ s.t. $[C'+r]_k - C' - r \ne 0$ (what these values of $C'$ are depends on the value of $r$). 
Likewise, and a bit simpler, the $\ell\times \ell$ component of $\Sigma_3'$ resembles $P_\ell$, but has an $x_4$-dependent phase in elements  with $[C-1]_\ell - C +1 \ne 0$.  Noting that $[C-1]_\ell - C - 1 = 0$   for $2 \le C \le \ell$, while, for $C=1$, it equals $[1-1]_\ell= \ell$, we find that (det$\Sigma_3'$)$^{1/N}$$=$$e^{- i 2 \pi {N q_3 + 1 \over N} {x_4 \over L_4}}$,  precisely the $U(1)$ transition function $\omega_3$ from (\ref{u1transition}). Thus, it is   the  extra $x_{2,4}$-dependent phases ``scattered'' inside the permutation matrices that ensure that the $U(1)$ transition functions extracted from (\ref{newgauge13}) remain the same as in (\ref{u1transition}).

Finally, let us find the gauge potentials   obtained after acting with (\ref{gauge12}). To this end, we let   $(\ln Q_\ell)_C$ denote  the $C$-th eigenvalue of the   matrix (likewise for $Q_k$), explicitly
$
(\ln Q_\ell)_C = i {\pi \over \ell} (2C - 1 - \ell)~.
$
This is relevant because $- i g \partial_{4} g^{-1} = - {i \over L_{4 }} \ln Q_{\ell }$ (and similar for $-i g \partial_2 g^{-1}$ with $\ell \rightarrow k$).
Thus, we can find the gauge transform (\ref{gauge12}) of (\ref{UNbackground2}, \ref{UNbackground3}). It is clear that   ${\cal A}_1'$ and ${\cal A}_3'$ remain equal to ${\cal A}_1$ and ${\cal A}_3$ after acting with $g(x)$ on the latter. 
The ${\cal A}_2'$ and ${\cal A}_4'$  are slightly different, as we see below.
In the new gauge, the $k \times k$ components of the background are, recalling from (\ref{SUNmoduli}) that $\phi_{\mu \; C'} = \phi_{\mu \; [C'-r]_k}$ and $\tilde\phi_\mu = {1 \over k} \sum\limits_{C'=1}^k \phi_{\mu \; C'}$,\begin{eqnarray}
\label{A1primeK}
{\cal A}'_{1, C'D'} &=& 2 \pi \;\delta_{C'D'}(z_1 -   \ell \phi_{1 \; C'})  \nonumber \\
{\cal A}'_{2, C'D'}  &=& {2 \pi  \over L_2} \;\delta_{C'D'}\left(  (q_1 - {r \over k}) {x_1 \over L_1 }  + L_2( z_2 -  \ell \phi_{2 \; C'}) +    {1 \over k}(C' - {1 +k \over 2})\right) \nonumber \\
 {\cal A}'_{3, C'D'} &=&  2\pi\;\delta_{C'D'}(z_3 -  \ell \phi_{3 \; C'}  )  \\
 {\cal A}'_{4, C'D'} &=& {2\pi \over L_4} \;\delta_{C'D'}(  q_3 {x_3 \over L_3 }+ L_4(z_4 - \ell \phi_{4 \; C'})) \nonumber
 \end{eqnarray}
while the $\ell\times \ell$ components read:
 \begin{eqnarray}\label{A1primeL}
{\cal A}'_{1, CD} &=& 2 \pi   \;\delta_{CD}(z_1+  k \tilde\phi_1)  \nonumber \\
{\cal A}'_{2, CD}  &=& {2 \pi\over L_2}\;  \delta_{CD}(  q_1 {x_1 \over L_1 } + L_2(  z_2+k \tilde\phi_2  ) )\nonumber \\
 {\cal A}'_{3, CD} &=&  2\pi \; \delta_{CD}(z_3+  k \tilde\phi_3)  \\
{\cal A}'_{4, CD} &=& {2\pi \over L_4}\;  \delta_{CD}\left( {x_3 \over L_3 } (q_3 + {1 \over \ell}) +L_4( z_4 +    k \tilde\phi_4)   + {1 \over \ell}(C  - {1 +\ell \over 2})\right) \nonumber 
\end{eqnarray}
We notice that the difference between (\ref{A1primeK}, \ref{A1primeL}) and (\ref{UNbackground2}, \ref{UNbackground3}) is only in the $x_2$ and $x_4$ components, in particular, in the terms ${1 \over k} (C' +...)$ and ${1 \over \ell}(C + ...)$ appearing on the r.h.s. after the gauge transformation (\ref{gauge12}). Needless to say, these terms are crucial to ensure that ${\cal A}'$ obey the proper boundary conditions with transition functions $\Sigma_\mu'$. They are also essential in understanding the space-time structure of the different wrapped stacks of branes on $\tilde \T^4$.

\subsection{The branes on the dual $\tilde\T^4$}
\label{sec:branesdual}

Now, given the gauge background, eqns.~(\ref{A1primeK}, \ref{A1primeL}), on the $D_{p+4}$ branes wrapped on $\T^4$, with transition functions (\ref{newgauge13}, \ref{newgauge24}), now conveniently trivial in $x_2$ and $x_4$,
 we perform $T$-duality in $x_2$ and $x_4$. We use the coordinates $y_\mu$ on the dual $\tilde \T^4$.  We recall that $y_1 = x_1$ and $y_3 = x_3$ (thus, $\hat L_1 = L_1, \hat L_3 = L_3$, while $\hat L_2 = 4 \pi^2 \alpha'/ L_2$, $\hat L_4 = 4 \pi^2 \alpha'/ L_4$, as per (\ref{tdual1})).
The $T$-dual (in $x_{2,4})$ of the $D_{p+4}$ background consists of two kinds of $D_{p+2}$ branes wrapped on cycles on the dual $\tilde \T^4$. The relation between the coordinates of the $D_{p+2}$ branes in $y_2$ and $y_4$ and the backgrounds ${\cal A}_{2,4}'$ are:\footnote{See \cite{Taylor:1996ik} for a derivation of (\ref{Ydual}) and  a discussion of $T$-duality from the point of view of the low-energy worldvolume field theory.} 
\begin{eqnarray}\label{Ydual}
\hat Y_2 &=& (2 \pi \alpha')(i \hat\partial_2 + {\cal{A}}_2'), \nonumber \\
\hat Y_4 &=& (2 \pi \alpha')(i \hat\partial_4 + {\cal{A}}_4'). 
\end{eqnarray}
 The hats over the $\tilde\T^4$ coordinates $\hat Y_\mu$ and over the partial derivatives serve to remind us that the $D_{p+2}$ branes live on the infinite cover of the dual torus, i.e.~the coordinates $\hat Y_2$, $\hat Y_4$ are infinite dimensional matrices, such that, e.g. $(\hat Y_2)_{n_2; n_4}$ describes (the lightest modes of) open strings stretched between a chosen brane and its $n_2$-th and $n_4$-th  images in the $y_2$ and $y_4$ directions (we can call these ``winding open strings,'' dual to Kaluza-Klein modes in $x_2, x_4$). 
 In what follows we shall be interested in the $y_2 = (Y_2)_{0;0}$ component (noting that $\hat \partial_{0}=0$ \cite{Taylor:1996ik}), i.e.~we ignore winding strings in our  study of the moduli space.
 
Thus, for the $k \times k$ components we find for the wrapped $D_{p+2}$-branes coordinates on the dual $\tilde\T^4$, recalling (\ref{tdual1}):
\begin{eqnarray}\label{hatYk}
{ y_{2 \; C'D'} \over \hat L_2} &=& \delta_{C'D'} \left[(q_1 - {r \over k}) {y_1 \over \hat L_1} + L_2(z_2 - \ell \phi_{2 C'})  +    {1 \over k}(C' - {1 +k \over 2})\right], \nonumber \\
{ y_{4 \; C'D'} \over \hat L_4} &=& \delta_{C'D'} \left[q_3  {y_3 \over \hat L_3} + L_4(z_4 - \ell \phi_{4 C'})\right],
\end{eqnarray}
with the  worldvolume Wilson lines inherited from (\ref{A1primeK})
\begin{eqnarray}
{\hat L_1 {\cal A}'_{1 \; C'D'} \over 2 \pi} = \delta_{C'D'} \hat L_1 (z_1 - \ell \phi_{1 C'})~, ~~
{\hat L_3 {\cal A}'_{3 \; C'D'} \over 2 \pi} = \delta_{C'D'} \hat L_3 (z_3 - \ell \phi_{3 C'})~.
\end{eqnarray}

{\flushleft{S}}imilarly, for the $\ell \times \ell$ components we find for the wrapped $D_{p+2}$-branes on $\tilde\T^4$:
\begin{eqnarray}\label{hatYell}
{y_{2 \; CD} \over \hat L_2} &=& \delta_{CD} \left[q_1  {y_1 \over \hat L_1} + L_2(z_2 + k \tilde\phi_{2}) \right], \nonumber \\
{y_{4 \; CD} \over \hat L_4} &=& \delta_{CD} \left[(q_3 + {1 \over \ell})  {y_3 \over \hat L_3} + L_4(z_4 + k \tilde\phi_{4})  + {1 \over \ell}(C  - {1 +\ell \over 2})\right],
\end{eqnarray}
with Wilson lines from (\ref{A1primeL}):
\begin{eqnarray}
{\hat L_1 {\cal A}'_{1 \; CD} \over 2 \pi} = \delta_{CD} \hat L_1 (z_1 + k \tilde\phi_{1})~, ~~{\hat L_3 {\cal A}'_{3 \; CD} \over 2 \pi} = \delta_{C'D'} \hat L_3 (z_3 + k\tilde\phi_{3})~.
\end{eqnarray}
As we will see in Section \ref{sec:numberofintersect} these equations give rise to a system of $D$-branes which interconnect in a somewhat intricate way as they wind around the torus. 
However, before we continue with a study of the windings and  intersection numbers of the various branes, we first consider the supersymmetry condition on the covering space of $\tilde\T^4$. 

\begin{figure}[h] 
   \centering
   \includegraphics[width=6in]{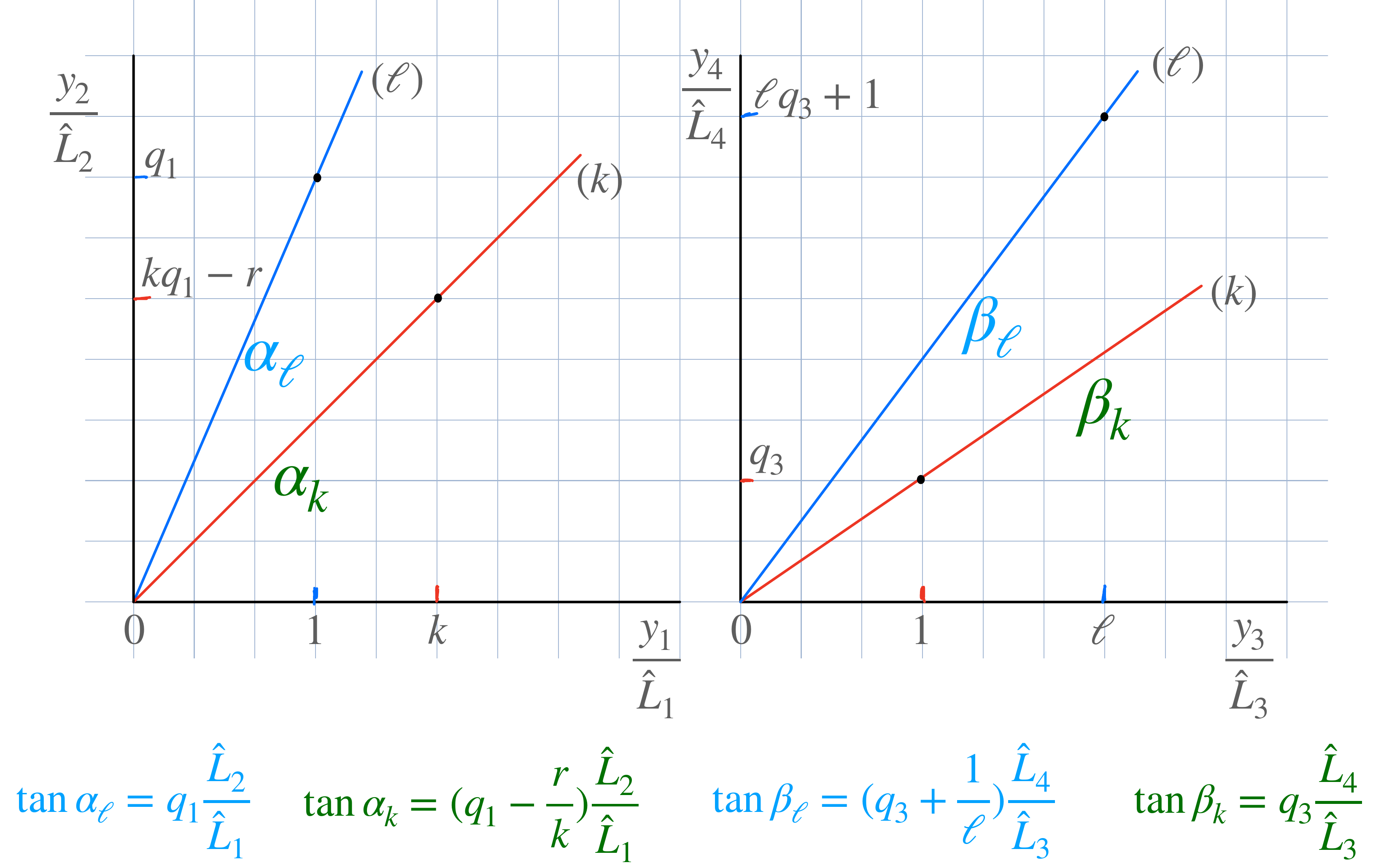} 
   \caption{The two tilted $D_{p+2}$-branes stacks, denoted by  $(k)$ and $(\ell)$, on the covering space of $\tilde\T^4$, plotted with  coordinates divided by the corresponding  period $\hat L_\mu$. Each $D_{p+2}$-brane stack can be thought as being initially in the $13$ plane,  then rotated on angle $\alpha$ into the $12$ plane and angle $\beta$ into the $34$ plane, as shown on the Figure. The angles are defined to be between the corresponding brane and the horizontal axis. The slopes follow from (\ref{hatYk}, \ref{hatYell}), with $q_{1,3}$  assumed nonnegative and $r$ positive (the grid in the background is not drawn to scale  and the black dots are meant to help visualize the slopes).  The supersymmetry condition \cite{Berkooz:1996km} on the covering space requires that there exist a complex structure such that the relative rotation between the two stacks is in $SU(2) \in SO(4)$, i.e. $\alpha_{\ell} - \alpha_k = \beta_{\ell} - \beta_k$ or eqn.~(\ref{bps0}). Explicitly, this corresponds to choosing $z_1 = y_1' + i y_2'$, $z_2 = y_4' + i y_3'$. The primed coordinates axes  $y_1', y_3'$ are chosen along the $(k)$ brane, while  $y_2'$, $y_4'$ are perpendicular to it; the primed axes are not shown to not crowd the Figure. The $SU(2)$ rotation of the $(k)$ stack into the $(\ell)$ stack is $(z_1, z_2)$ $\rightarrow (e^{i x} z_1, e^{-i x} z_2)$ with $x= \alpha_\ell - \beta_\ell = \alpha_k - \beta_k$. When the angles are chosen to preserve supersymmetry, there is no open string tachyon, but massless states of the $k$-$\ell$ strings, localized at the  intersection. These comprise a hypermultiplet, in 4d ${\cal N}=2$ language, bifundamental under the gauge groups on the worldvolumes of the intersecting branes.}
   \label{brane4}
\end{figure}

\subsection{The BPS conditions from  branes-at-angle on the covering space} 
\label{sec:bps}

On Figure \ref{brane4}, we plot  the $(k)$ and $(\ell)$ $D_{p+2}$ branes, embedded in the covering space, in a manner described by eqns.~(\ref{hatYk}, \ref{hatYell}). 
One set of branes ($k$, the red one) has worldvolume along the lines ${y_2} = (q_1 - {r \over k}) {\hat L_2 \over \hat L_1}  y_1$, ${y_4 } = q_3 {\hat L_4 \over \hat L_3} y_3$, while the other ($\ell$, the blue one) is along {${y_2} = q_1  {\hat L_2 \over \hat L_1}  y_1$, ${y_4 } = (q_3 + {1 \over \ell}) {\hat L_4 \over \hat L_3} y_3$. Clearly, the covering space angles are determined by $q_{1,3}, r, k, \ell$ and the ratios of sides of the dual $\tilde \T^4$.
On the picture, we have taken the values of the moduli, or brane positions, such that the branes intersect at the origin. 

As is well known \cite{Berkooz:1996km} and reviewed in  the caption of Figure \ref{brane4},  the intersecting brane BPS condition is 
\begin{eqnarray}\label{bps0}
\alpha_\ell - \beta_\ell = \alpha_k - \beta_k~,
\end{eqnarray}
with the angles  defined in the Figure. 
This condition implies,\footnote{\label{foot:tangent}Recalling $\tan(a-b) = (\tan a - \tan b)/(1 + \tan a \tan b)$.}
\begin{eqnarray}\label{bps1}
 {r \ell \over k} {\hat L_2 \hat L_3 \over \hat L_1 \hat L_4 } =  { 1 + q_1 (q_1 - {r \over k}) {\hat L_2^2 \over \hat L_1^2} \over  1 + q_3 (q_3 + {1 \over \ell}) {\hat L_4^2 \over \hat L_3^2}} \implies {r \ell \over k} {L_3 L_4 \over L_1 L_2} = { 1 + q_1 (q_1 - {r \over k}) {k \over r \ell} {16 \pi^4 \alpha'^2 \over V_{\T^4}} \over  1 + q_3 (q_3 + {1 \over \ell}) {r \ell \over k} {16 \pi^4 \alpha'^2 \over V_{\T^4}}},
\end{eqnarray}
where we used the $T$-duality relations (\ref{tdual1}) and denoted the volume of $\T^4$ by $V_{\T^4} = L_1 L_2 L_3 L_4$. 

We begin by  noting  that with zero $D_p$-brane charge,\footnote{On the worldvolume of the $D_{p+4}$ brane, a background of nonzero second Chern character (\ref{secondchern}) on the $\T^4$ carries $D_p$-brane (Ramond-Ramond) charge. Likewise, the nonvanishing first Chern characters, given by eqn.~(\ref{fluxU1}) times $N$, or $U(1)$ fluxes, give rise to $D_{p+2}$-brane charges.} i.e. for $q_1=q_3=0$,  using the $T$-duality relations (\ref{tdual1}), eqn.~(\ref{bps1}) is nothing but the $SU(N)$ self-duality condition obtained earlier in our field-theory analysis (\ref{SUNBPS}): 
\begin{equation}\label{branebps1}
 {r \ell \over k} {\hat L_2 \hat L_3 \over \hat L_1 \hat L_4 } =  {r \ell \over k} {L_3 L_4 \over L_1 L_2}=1 .
\end{equation}
Thus, with zero $D_p$-brane charge, or vanishing $U(N)$ 2nd Chern character (\ref{secondchern}), the supersymmetry condition is satisfied by imposing $SU(N)$ self-duality. It constrains the ratio of   the torus periods (or the complex structure---the ratio of lengths is similar to the  $\tau$ parameter of a two-torus).

If the $D_p$-brane charge (\ref{secondchern}) is nonzero, we also impose (\ref{branebps1}), the constraint on the complex structure  of $\T^4$.\footnote{This matches the field theory finding, where $U(N)$ self duality imposes $SU(N)$ self duality. Let us note that, mathematically, eqn.~(\ref{bps1}) could be solved, for arbitrary $L_3 L_4 \over L_1 L_2$  (assuming $V_{\T^4}$ comes out positive) by solving for $V_{\T^4}/\alpha'^{2}$ in terms of the period ratio and $q_{1,3}, r, \ell, k$. This, however, would fix the ratio of volume to $\alpha'^2$ (a parameter not existing in field theory) in terms of the complex structure and the  integer valued data,  and we discard it.}
From the r.h.s. of  (\ref{bps1}) it then follows that eqn.~(\ref{bps1}) reduces to the requirement
\begin{eqnarray}\label{bps2}
  { 1 + q_1 (q_1 - {r \over k}) {k \over r \ell} {16 \pi^4 \alpha'^2 \over V_{\T^4}} \over  1 + q_3 (q_3 + {1 \over \ell}) {r \ell \over k} {16 \pi^4 \alpha'^2 \over V_{\T^4}}} = 1~.
\end{eqnarray}
As supersymmetry should hold for a continuous range of volumes, it must be, therefore, that
\begin{eqnarray}\label{bps3}
k q_1 (k q_1 - r)   =  r    \ell  q_3 (r \ell q_3 + r)~\implies ~k q_1 = {r  \ell q_3+ r } .
\end{eqnarray} As we indicated above,  the solution of eqn.~(\ref{bps2}) is exactly the $U(N)$ self-duality condition (\ref{UNBPS}), with both $q_{1,3}$ positive integers. We conclude that with nonzero $D_p$-brane charge, the supersymmetry condition on the covering space is equivalent to  $U(N)$ self-duality.

Let us also go back to the geometric interpretation of the solutions of the supersymmetry conditions, (\ref{bps1}) and (\ref{bps3}). When $q_1 =q_3=0$, we find, from (\ref{bps1}) that $\alpha_\ell = \beta_k =0$ and $\beta_\ell = - \alpha_k$ and thus (\ref{bps0}) is obeyed. On the other hand, from (\ref{bps3}), we find, again using (\ref{bps1}), that $\alpha_\ell = \beta_\ell$ and $\alpha_k = \beta_k$, also obeying (\ref{bps0}).\footnote{One can study the possibility of having more than two stacks of branes wrapped on $\tilde\T^4$ obeying the BPS conditions.  The $\beta_\ell = - \alpha_k$ and $\beta_k=- \alpha_\ell$  solution (the one obtained above with $q_1=q_3=0$) only allows two stacks of branes. However, the angle configuration $\alpha_{i} = \beta_i$   (obtained for $q_{1,3}\ne 0$ above, with $i= \{\ell, k\}$) allows for BPS configurations with more than two stacks by extending further the range of $i$. A detailed study of the generalization of 't Hooft's construction to  $N=k_1 + k_2 +... + k_p$ with $p>2$, $k_i \in \mathbb{N}$, is beyond our scope here.}

Upon quantizing the strings connecting the two intersecting stacks of branes, one finds that there is no open string tachyon when the supersymmetry condition is obeyed \cite{Berkooz:1996km}.  There are, instead, exactly massless open string states, localized at the brane  intersection. In terms of ${\cal N}=2$ 4d supersymmetry multiplets, they comprise a massless hypermultiplet transforming as a bifundamental under the gauge groups living on the worldvolumes of the two intersecting branes. 
As we  describe in the next Section, when the covering space picture is mapped to the $\tilde\T^4$ in a manner described by the equations of  Section \ref{sec:branesdual}, the branes interconnect in interesting ways and there is, for generic values of $r, k, \ell$, more than one  intersection point between the different stacks. There are massless string states localized at each  intersection. Their enumeration is crucial for the description of the moduli space.

\bigskip

\subsection{Intersection numbers of wrapped $D$-branes and Ramond-Ramond charges}
\label{sec:numberofintersect}

We now go back to the equations for the tilted $D_{p+2}$ branes on $\tilde \T^4$, given in eqns. (\ref{hatYk}) and (\ref{hatYell}), taking $q_1 = q_3=0$  for simplicity. For brevity, we also introduce the notation 
\begin{eqnarray}
g \equiv \text{gcd}(k,r) \label{gnotation},
\end{eqnarray}
which, as we recall from (\ref{SUNmoduli}), determines the number of independent $SU(N)$ moduli  for each direction of $\T^4$.
We begin equation (\ref{hatYk}) with $q_1=q_3=0$:  
\begin{eqnarray}\label{hatYk1}
{ y_{2 \; C'D'} \over \hat L_2} &=& \delta_{C'D'} \left[ - {r \over k}  {y_1 \over \hat L_1} + L_2(z_2 - \ell \phi_{2 C'})  +    {1 \over k}(C' - {1 +k \over 2})\right], \nonumber \\
{ y_{4 \; C'D'} \over \hat L_4} &=& \delta_{C'D'} \left[ L_4(z_4 - \ell \phi_{4 C'})\right], ~C', D' = 1,...,k.
\end{eqnarray} 
This describes a stack of $g$ branes---one brane for each independent value of $\phi_{\mu C'}$ from (\ref{SUNmoduli})---wrapping $-r/g$ times around $y_2$, $k/g$ times around $y_1$, at fixed position in $y_4$ (with the world volume  along $y_3$). 

To see this, consider  the $1$-$2$ plane equation, where the appearance of $C'$ on the r.h.s. shows that different branes ($C'$ and $[C'-r]_k$) reconnect  after winding around $y_1$. Consider the $y_2\over \hat L_2$ coordinate of the $C'$-th brane upon reaching $y_1=\hat L_1$:
\begin{eqnarray}\label{y11}
{y_{2 C' C'}\over \hat L_2}\vert_{y_1 = L_1}  = - {r \over k} + L_2 (z_2 - \ell \phi_{2 C'}) + {1\over k}(C' - {1+k \over 2})~.
\end{eqnarray} But this is the same, (mod $1$), as the  ${y_2 \over \hat L_2}$ coordinate, at $y_1=0$,  of the $[C'-r]_k$-th brane. Explicitly,   from (\ref{hatYk1}), 
\begin{eqnarray}\label{y22}
 {y_{2 [C'-r]_k [C'-r]_k}\over \hat L_2}\vert_{y_1 = 0} =  L_2 (z_2 - \ell \phi_{2 [C'-r]_k}) + {1 \over k}([C'-r]_k - {1 + k \over 2}).\end{eqnarray} We now recall, from (\ref{SUNmoduli}), that  $\phi_{2 C'} = \phi_{2 [C'-r]_k}$ and use ${[C'-r]_k \over k} = {C'\over k} - {r \over k} \; (\text{mod} 1)$, to conclude that, indeed (\ref{y11}) and (\ref{y22}) are equal (mod $1$). 

Thus, the trajectory of each of the $g$ branes in the covering of the $12$ plane is ${k \over g} {y_2 \over \hat L_2} = - {r \over g} {y_1 \over \hat L_1}$. The two coordinates are equal integers at the origin and at ${y_1\over \hat L_1} = {k\over g}$ and ${y_2\over \hat L_2} = -{r \over g}$. This implies, as stated after (\ref{hatYk1}), that the brane wraps around, as a one dimensional curve, $k/g$ times in $y_1$ and $-r/g$ times in $y_2$. 
Ultimately, this is a consequence of the fact that the transition function in the $y_1$ direction, $\Sigma_1'$ from (\ref{newgauge13}) involves (a modified version of) the permutation matrix $P_k^{-r}$. In the $34$ plane, on the other hand, the $g$ distinct $C'$ branes are parallel to $y_3$. 

Moving to the $C$ brane---and equations (\ref{hatYell}) with $q_1=q_3=0$,  giving rise to $\ell {y_4 \over \hat L_4} = {y_3 \over \hat L_3}$  (and $y_2 = 0$, up to constants) on the covering space, everything is exactly the same as in the analysis above---but the behaviour in the two planes is reversed: in the $34$ plane, there is a single $\ell$-brane wrapping $\ell$ times in $y_3$ (the $C$ brane connects with $[C-1]_\ell$  one upon traversing the $y_1$ direction), while in the $12$ plane, it is parallel to the $y_1$ axes, at fixed position in $y_2$.

From the above---and this is very important for our analysis of the moduli space---it follows that each of the $g$ ``$k$''-branes wrapping $-r/g$ times in the $y_1$ direction has $r/g$  intersections in the $12$ plane with the single $\ell$-brane, which is parallel to $y_1$. Each of the $r/g$  intersection points of a given ``$k$''-brane with the  ``$\ell$''-brane have different coordinates in the $12$ plane but the same coordinate in the $34$ plane.

To illustrate this, 
let us consider a concrete example and a picture.  On Figure \ref{brane5}, we show the brane configurations  in the $12$ and $34$ planes of $\tilde\T^4$, for $k=6, r=4$, $\ell = 2$. Thus, $P_k^{-r} =P_6^2$ for the values plotted, indicating that $C'=1$ connects with $C'=3$, $C'=3$ with $C'=5$, and $C'=5$ with $C'=1$, and similar for the even values of $C'$ (as discussed above, $C'$ reconnects with $[C'-4]_6$). This  leads to the picture of two parallel branes  (as $g=2$) wrapping $3$ times in $y_1$, $2$ times in $y_2$, as shown, as well as once in $y_3$ and with no wrapping in $y_4$.
Similarly, the interpretation of equation (\ref{hatYell}) is that there is a single $D$-brane wrapping once in $y_4$,  $\ell$ times in $y_3$,  without any wrapping in $y_2$ and a wrapping once in $y_1$.
Likewise, the $\ell$-tuple winding in $y_3$ is a consequence of the fact that the  transition function $\Sigma'_3$ in (\ref{newgauge13}) includes the permutation matrix $P_\ell$ ($=P_2$ as plotted). It is clear from the Figure that each of the $g$ (two) $k$-branes have $r/g$ (two)  intersections with the single $\ell$-brane.

{\flushleft{\bf Brane wrappings and Ramond-Ramond charges:}} A slightly different perspective on the wrapping of the two-dimensional worldvolume on the $\tilde\T^4$ is also useful. A map from a worldvolume two torus, $\T^2$, parameterized by $\sigma \in \R \; (\text{mod}\; 1)$ and $\tau \in \R \; (\text{mod}\; 1)$, to the $\tilde\T^4$, the latter parameterized by 
$t_\mu \equiv {y_\mu \over \hat L_\mu} \in \R \; (\text{mod}\; 1)$, is characterized by the six wrapping numbers $w_{\mu\nu} = - w_{\nu\mu}$ (the six non contractible two-planes form a basis of $H_2(\tilde \T^4, \Z)$):
\begin{eqnarray}\label{wrap1}
w_{\mu\nu} = \int\limits_{\T^2} d \sigma  d \tau \left({\partial t_\mu \over \partial \sigma} {\partial t_\nu \over \partial \tau} - {\partial t_\nu \over \partial \sigma} {\partial t_\mu \over \partial \tau}\right)~.
\end{eqnarray}
Let us denote the numbers $w_{\mu\nu}$ for the map (\ref{hatYk}) with $q_1 =q_3=0$,  by $w_{\mu\nu}^{(k)}$. Parameterizing $t_1 = {k \over g} \sigma$, $t_2 = - {r \over g}\sigma$, $t_3 = \tau$ and $t_4 = 0$, omitting all constant terms, we find\footnote{To avoid any confusion, recall eqn.~(\ref{gnotation}), showing that all winding numbers below are integers.}
\begin{eqnarray}\label{wrap2}
w_{13}^{(k)} = {k \over g}, \; w_{23}^{(k)} = - {r \over g}, \; w_{12}^{(k)} = w_{14}^{(k)} = w_{24}^{(k)} = w_{34}^{(k)} = 0  ~ ~(\text{$\times$\; $g$ branes}),
\end{eqnarray}
where we indicated that there are $g$ branes with the above winding numbers.
For the map (\ref{hatYell}), we denote the winding numbers by $w_{\mu\nu}^{(\ell)}$, parameterizing $t_1 = \sigma$, $t_2 =0$, $t_3 = \ell \tau$ and $t_4 = \tau$, the winding numbers are
\begin{eqnarray}\label{wrap3}
w_{13}^{(\ell)} = \ell , \; w_{14}^{(\ell)} = 1, \; w_{12}^{(\ell)} = w_{23}^{(\ell)} = w_{24}^{(\ell)} = w_{34}^{(\ell)} = 0  ~~ (\text{$\times$\; $1$ brane}),
\end{eqnarray}
where we indicate that there is a single brane with these wrapping numbers. 

These wrapping numbers translate to charges under the Ramond-Ramond (RR) fields sourced by the branes. The stack of $g$ branes with wrapping numbers (\ref{wrap2}) has $D_2$ brane\footnote{For this discussion, for definiteness, we take $p=0$.} charges $g \times w_{13}^{(k)} = k$ in the $13$-plane and $g \times w_{23}^{(k)} = - r$ in the $23$ plane, while the single $\ell$-brane (\ref{wrap3}) has $D_2$ brane charge $w_{13}^{(\ell)} = \ell$ in the $13$ plane and $w_{14}^{(\ell)} = 1$ in the $14$ plane. Thus the total RR $D_2$ brane charge in the $13$-plane is $N$, while it is $-r$ in the $23$ plane and $1$ in the $14$ plane. Now we recall how RR charges transform under $T$ duality. Upon $T$-duality in the $y_2$ and $y_4$ directions, the $D_2$-brane charge in the $13$ plane becomes $D_4$-brane charge (the brane now wraps the entire $\T^4$). Similarly, the winding      $w_{23}$   gives $N$ times the $U(1)$ flux in the $12$ plane of $\T^4$ (eqn.~(\ref{fluxU1}) with $q_1 = q_3=0$ times $N$) and   $w_{14}$ similarly yields  $N$ times the $U(1)$ flux in the  $34$ plane of $\T^4$.\footnote{This follows from the transformations of RR fields under $T$-duality \cite{Bergshoeff:1995as}, which can be elegantly written using an index formula; see the first equation in \cite{Hori:1999me}. We skip  the straightforward details and just state  the result.}

We stress that the point of the above paragraph  is to emphasize that the appearance of two stacks, one of $g$ parallel ``$k$''-branes and the other of a single ``$\ell$''-brane  with winding numbers (\ref{wrap2}, \ref{wrap3})---which, at first sight, might seem strange in a $U(N)$ theory---is, in fact, consistent with the various RR charges.

Let us pause for a moment and consider the situation without flux. 
The wrapping numbers and brane multiplicities from (\ref{wrap2}, \ref{wrap3}) are to be contrasted with those for the case without flux, where the only nonzero windings\footnote{With trivial transition functions (a multiply-wrapped $D$-brane  differs from many $D$-branes each wrapped once by the background Wilson line---equivalently, the transition function \cite{Polchinski:1996fm}). } are:
\begin{eqnarray}\label{wrap4}
w_{13}^{(k)} = 1 \; (\text{$\times$\; $k$ branes}),  \;\; w_{13}^{(\ell)} = 1  \; (\text{$\times$\; $\ell$ branes}).
\end{eqnarray}
As per the discussion of the previous two paragraphs, the  windings (\ref{wrap4}) correspond, on the dual $\T^4$, to total $D_4$-brane charge $N$, without any $D_2$-brane charges.

{\flushleft{\bf D-brane embedding and coordinate conventions:}} The configuration without flux described by (\ref{wrap4}) can be thought of as  $T$-dual, in   $x_2$ and $x_4$, as in (\ref{tdual1}), of a configuration of $N$ $D_7$ branes\footnote{\label{footnoteatwill}We stress that introducing the extra noncompact directions of the brane worldvolume is only done for convenience---it allows us to use 4d SYM  intuition to find the equations describing the moduli space. Replacing the $D_0$-$D_{4}$ system with, e.g. a $T$-dual $D_3$-$D_7$ one, is often done in the literature on the relation of ADHM to D-branes; see \cite{Tong:2005un}.}  wrapped on $\T^4$. 
For concreteness, we take the $D_7$ branes to have worldvolume along $x_0, x_1, x_2, x_3, x_4, x_5, x_6, x_7$ with $x_1,...,x_4$ being the $\T^4$ directions. 
The $T$-dual configuration on $\tilde\T^4$ is a configuration of   $N$ $D_5$ branes, with two of their worldvolume dimensions  wrapped  in the $13$ plane of $\tilde\T^4$ with unit winding, as in (\ref{wrap4}). The worldvolume directions of the $D_5$ branes are  $y_0, y_1, y_3, y_5, y_6, y_7$. They are wrapped in $y_1, y_3$ and  are localized in $y_2, y_4$ ($y_1,...,y_4$ are $\tilde\T^4$ coordinates). Their four-dimensional 
 extended worldvolume is in $y_0, y_5, y_6, y_7$.
 The long-distance theory in the four noncompact directions of the $D_5$ brane worldvolume  is ${\cal N}=4$ 4d $U(N)$ SYM. The six real adjoint scalars  of ${\cal N}=4$ SYM are   the separations between the branes in the $y_2, y_4$ directions inside $\tilde\T^4$, their separations in the two noncompact $y_8, y_9$ directions not in the $D_5$ worldvolume, and the Wilson lines in the $y_1, y_3$  compact directions of the worldvolume.
 
We now turn on flux on the worldvolume of the $D_7$ branes   wrapped on $\T^4$. Their 
worldvolume theory is  a compactified $16$ supercharge theory, but now with worldvolume fluxes turned on, as per (\ref{A1primeK}, \ref{A1primeL}). The flux background preserves $8$ supercharges for a torus whose sides satisfy (\ref{branebps1}), which is what we assume throughout.
The $T$-dual configuration with  $\tilde\T^4$ wrapping numbers  (\ref{wrap2}, \ref{wrap3}) is then one of wrapped intersecting $D_5$ branes.

 \begin{figure}[h] 
   \centering
   \includegraphics[width=6in]{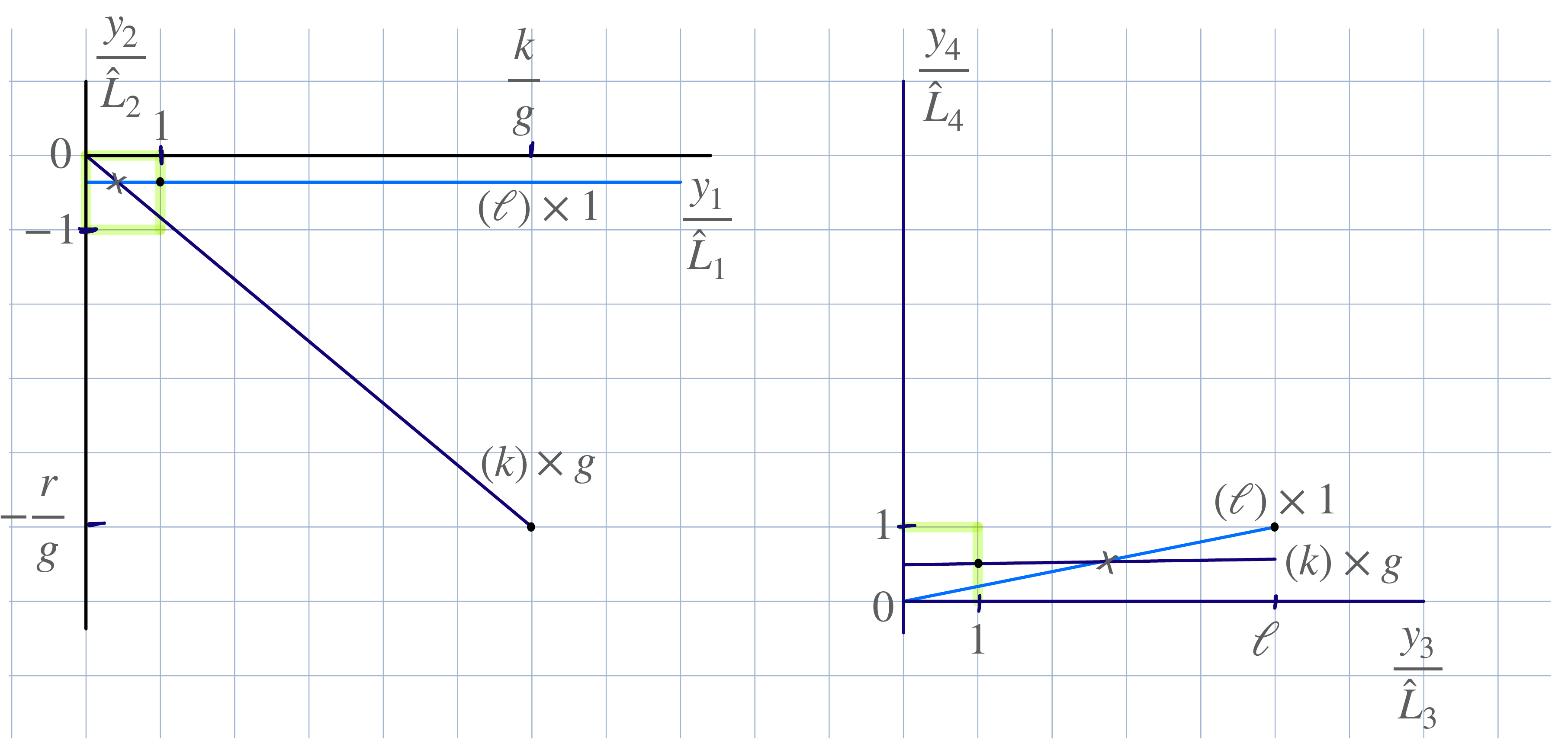} 
   \caption{A summary of   brane configuration dual to an instanton with  $Q={r \over N}$, $r \in {\mathbb{N}}$, $g\equiv{\rm{gcd}}(k,r)$, $N$$=$$k+ \ell$. The fundamental domain of $\tilde\T^4$ is indicated by the green unit square, taken to coincide with the background square grid. There is a single ``$\ell$''-brane, parallel to $y_1$ with unit winding in $y_1$, and winding $\ell$ times around $y_3$ and once in $y_4$. There are $g$ parallel ``$k$''-branes (all shown on top of each other to avoid overcrowding the picture), winding $k/g$ times in $y_1$,  $-r/g$ times in $y_2$, and parallel to and winding once around  $y_3$. The  intersections between the ``$k$''- and ``$\ell$''- branes are shown by a cross. Upon bringing the ``$k$'' brane into the fundamental domain in the $12$ plane, there are $r/g$ distinct  intersection points with the ``$\ell$''-brane (equivalently, these are the covering-space  intersections between the dark blue brane and the images of the light blue brane  translated by a period in the $y_2$ direction on the covering space).
In contrast, one observes that upon bringing the ``$\ell$''-brane into the fundamental domain in the $34$ plane, there is  only a single  intersection point with the ``$k$''-brane.}
   \label{brane6}
\end{figure}
Here we focus on the 4d long-distance $D_5$-brane worldvolume theory. For generic values of the moduli, a $U(1)_\ell \times U(1)^g = U(1)_\ell  \times \prod_{i=1}^g U(1)_i$ theory with ${\cal N}=2$ 4d supersymmetry. The first $U(1)$ factor lives on the ``$\ell$''-brane and the rest---on the $g$ ``$k$''-branes. Our interest is to describe the moduli space of the theory on the worldvolume of the stack of $g$ ``$k$''-branes and the ``$\ell$''-brane intersecting on the $\tilde\T^4$. As described above, we denote the $T$-dual space-time dimensions as $y_0, y_1, \ldots, y_9$, where the    $\tilde\T^4$ coordinates are $y_1, y_2, y_3, y_4$, as above. The $D_5$ branes have four noncompact worldvolume coordinates $y_0, y_5, y_6, y_7$, are wrapped on two-cycles in $\tilde\T^4$, as per (\ref{wrap2}, \ref{wrap3}), and are localized in $y_8$ and $y_9$.

{\flushleft{ \bf A pictorial summary of the wrapped intersecting brane configuration for general $r, k, g$ ($q_{1,3}=0$):}} The winding intersecting brane configuration for general $r, g, k$  is shown on Figure \ref{brane6}, on the covering space of $\tilde{\T}^4$. The visualization of the number of  intersection points is discussed in the caption.

 \subsection{The moduli space as the Higgs branch of a $4 d$ ${\cal N}=2$ supersymmetric theory}
  
\label{sec:u1branch}

Let us  describe the matter content of the long distance 4d theory using 4d ${\cal N}=1$ supersymmetry notation. Each  $U(1)$ factor has an ${\cal N}=1$ vector multiplet and an adjoint chiral multiplet (for a $U(1)$ group, the latter is neutral). The chiral adjoint  in the ${\cal N}=2$ vector supermultiplet describes the position of the brane in the noncompact directions orthogonal to the $\tilde\T^4$---as described above, there are two such noncompact directions, $y_8, y_9$. We call the chiral adjoints $\phi_3, \phi_{3 i}, i=1,...,g$, for the respective $U(1)$ factors.

 In addition there is an adjoint hypermultiplet (two ${\cal N}=1$ chiral multiplets), whose complex scalar components we denote $\phi_1, \phi_2$ for the first $U(1)_\ell$ factor and $\phi_1^i, \phi_2^i$, $i=1,...,g$ for the $U(1)_i$. For each gauge group, the four real components of $\phi_1, \phi_2$ describe the positions of the corresponding brane inside $\tilde\T^4$ and the gauge-field Wilson lines along the $2$-cycles  the brane is  wrapped on. Finally, for each $i = 1,...,g$ there are $r\over g$ hypermultiplets, arising from the massless excitations of the strings connecting the corresponding intersecting branes. There is one hypermultiplet localized at each of the $r/g$  intersection points between each ``$k$''-brane and the ``$\ell$''-brane. Each hypermultiplet consists of two ${\cal N}=1$ chiral multiplets. Their  scalar components are $q_i^a, \tilde{q}_a^i$, $a = 1,..., {r \over g}$. Each of the $r/g$ hypermultiplets  is charged under $U(1)_\ell \times U(1)_i$, where we recall that $i=1,...,g$. The fields and their charges under  $U(1)_\ell \times U(1)_1 \times ... \times U(1)_g$ are:
\begin{eqnarray}\label{mattercontent}
\begin{array}{c|c|c|c|c|c|c|c} \text{field}&U(1)_\ell \; \text{charge} &U(1)_1 \; \text{charge}  &... &U(1)_i  \; \text{charge} &... & U(1)_g  \; \text{charge} & \text{multiplicity} \cr \hline
q_1^a& 1& - 1& 0 & 0& 0 &0 & a=1,..., r/g \cr
\tilde{q}_a^1& -1 & 1& 0& 0& 0& 0& a=1,..., r/g\cr
\ldots & & & & & & & a=1,..., r/g\cr
q_i^a& 1& 0&0 & -1& 0&0 & a=1,..., r/g \cr
\tilde{q}_a^i & -1&0 &0 & 1&0 &0 & a=1,..., r/g\cr
\ldots & & & & & & & a=1,..., r/g\cr
q_g^a& 1&0 &0 &0 &0 &-1 & a=1,..., r/g \cr
\tilde{q}_a^g& -1&0 &0 &0 &0 &1 & a=1,..., r/g\cr
\phi_A & 0& 0& 0& 0& 0& 0& A=1,2,3  \cr
\phi_A^i& 0& 0&0 &0 &0 &0 & A=1,2,3; i = 1,...,g. \cr
\end{array}
\end{eqnarray}
Note that the hypers are $q_i^a, \tilde{q}_a^i$ neutral under the diagonal $U(1)$, i.e. do not couple to the overall ``center of mass'' of the branes. The chiral adjoints in the ${\cal N}=2$ vector multiplet are $(\phi_3, \phi_{3 i})$ and the chiral adjoint hypermultiplets are $\phi_1, \phi_2, \phi_{1 i}, \phi_{2 i}$, $i=1,...,g$, all of which are neutral under the $U(1)$'s.

{\flushleft{B}}efore we continue, let us  make the following comments on the scope (and limitation) of our study of the moduli space:
\begin{enumerate}
\item The Coulomb branch of the ${\cal N}=2$ theory described above corresponds to separating the two stacks of wrapped $D_{5}$ branes in the $y_8$ and $y_{9}$ directions orthogonal to the $\tilde{\T}^4$. It thus gives masses to the bifundamental hypermultiplets living at the  intersections. This causes additional breaking of the $T$-dual gauge group by these vector multiplet adjoint vevs (this is especially clear in the enhanced symmetry point, see the discussion further below) which has  nothing to do with the pure gauge theory background of interest (\ref{the full abelian bck general r}). Thus, the Coulomb branch does not describe self-dual deformations of the constant flux background.  On the other hand, 
the Higgs branch of the worldvolume theory moduli space  corresponds to the deformations of  the constant flux background that do not change the action and does not separate the branes in $y_{8,9}$.
\item Our focus on the scalar potential of the 4d EFT  suffices to explore the moduli space locally and to find its dimension   for general $r, k$---assuming it has no disconnected components. However, exploring its global structure or finding the explicit instanton background---especially for cases when the bifundamental hypers can have nonzero vevs (e.g. for $g \ne r$, see below)---requires resolving structures inside $\tilde\T^4$ and going beyond the 4d worldvolume EFT. This  is clear from the fact that our 4d ${\cal N}=2$ EFT with superpotential  given in (\ref{superpotential1}) below, ignores the compact nature of the fields $\phi_{1}, \phi_2$, etc., describing the moduli space.  This compact nature was crucial for describing its global structure---in the gcd$(k,r)=r$ case where it is understood in QFT (described at the end of Section \ref{sec:missing}). 

Despite the limitations of the EFT approach, a focus on the 4d EFT allows us to use our understanding of ${\cal N}=2$ supersymmetric Higgs branches. This will permit us to reproduce, in only a few lines, the results on the parameterization of the moduli space,  obtained in QFT after the lengthy and laborious calculations of \cite{Anber:2023sjn,Anber:2025yub}.
\end{enumerate}
This being said, we focus on the 4d EFT in the $y_0, y_5, y_6, y_7$ noncompact directions of the $D_5$ branes worldvolume. The 
 superpotential of the $U(1)_\ell \times U(1)^g$ theory, with charged fields given in the table above,  is fixed by ${\cal N}=2$ supersymmetry:
\begin{eqnarray}\label{superpotential1}
W = \sum_{i=1}^{g} \phi_{3 i} \sum_{a = 1}^{r \over g} \tilde{q}_a ^i q_i^a + \phi_3 \sum_{i=1}^{g} \sum_{a = 1}^{r \over g} \tilde{q}_a^i q_i^a~,
\end{eqnarray}
where we absorb the inessential normalization into the fields $\phi_3, \phi_{3 i }$. We stress that the adjoint hypermultiplets $\phi_{1}, \phi_2, \phi_{1i}, \phi_{2i}$ do not appear in either the $F$-terms or the $D$-terms. The $F$-term conditions resulting from the variation of (\ref{superpotential1}) wrt the bifundamental hypermultiplets $q_i^a$ and $\tilde q_a^i$ are:
\begin{eqnarray}\label{fterm1}
(\phi_3 + \phi_{3 i}) \tilde{q}_i^a = 0, ~(\phi_3 + \phi_{3 i}) {q}_i^a = 0~.
\end{eqnarray}
To continue, as we already stressed, we are working in a setup where the two stacks of branes are not separated in the directions orthogonal to $\tilde\T^4$ (otherwise, the $SU(N)$ gauge group would be broken by the separation and not only by   the flux background). In other words, we are interested in the Higgs, not Coulomb branch of the $U(1)_\ell \times U(1)^g$ theory. Thus, we set $\phi_3 + \phi_{3 i}=0$ (ignoring the motion of the center of mass of the entire configuration in $y_{8,9}$). The remaining  $F$-term conditions are from the variation of $W$ wrt $\phi_{3i}, \phi_{3}$:
\begin{eqnarray}\label{fterm2}
\sum_{a=1}^{r\over g} \tilde{q}_a^i q_i^a = 0, ~i=1,...g;  ~ \sum_{i=1}^g \sum_{a=1}^{r\over g} \tilde{q}_a^i q_i^a = 0 ~~ (\text{$g$ complex constraints}),
\end{eqnarray}
while the $D$-term conditions are:\footnote{Note that there are no deformations in this setup that give rise to Fayet-Illiopoulos terms.}
\begin{eqnarray}\label{dterm2}
\sum_{a=1}^{r\over g} |\tilde{q}_{a}^i|^2 - |q_{i}^a|^2 = 0,  ~i=1,...g;~\sum_{i=1}^g \sum_{a=1}^{r\over g} |\tilde{q}_{a}^i|^2 - |q_{i}^a|^2 = 0 ~~ (\text{$g$ real constraints}).
\end{eqnarray}
We note that the second equations in both (\ref{fterm2}) and (\ref{dterm2}) follow from the first, which is a reflection of the fact that the hypermultiplets are uncharged under the diagonal $U(1)$.

Let us now count the dimension of the Higgs branch, or equivalently the dimension of the moduli space. The adjoint hypers $\phi_{1}, \phi_2, \phi_{1 i}, \phi_{2 i}$ give rise to $4 g + 4$ moduli.\footnote{We stress that these are nothing but the moduli present in (\ref{hatYk}, \ref{hatYell}), renamed.} The bifundamental hypers $q_{i}^a, \tilde{q}_{a}^i$ ($i=1,...,g; a=1,...,{r \over g}$) have $2  \times g \times {r \over g} = 2 r$ complex or $4 r$ real components. These are subject to the $3 g$ real constraints from the $F$-term and $D$-term conditions (\ref{fterm2}, \ref{dterm2}). In addition, the nontrivially acting gauge factor  ($U(1)^g$)  should be modded out, increasing the number of constraints to a total of $4 g$ real constraints. 
Thus, adding the number of fields and subtracting the number of constraints, we find that the real dimension of the Higgs branch is:
\begin{eqnarray}\label{dimensionmoduli}
\text{dim of Higgs branch} = \underbrace{4 g + 4}_{\text{adjoint hypers}} + \underbrace{4r}_{\text{bifundamental hypers}} - \underbrace{4 g}_{D-, F-, {\text{gauge constraints}}} = 4r + 4.\nonumber \\
\end{eqnarray} 
We note that the factor of $4r - 4g$ is the dimension of the bifundamental hypermultiplet part of the Higgs branch.
Both eqns.~(\ref{dimensionmoduli}) and (\ref{fterm2}, \ref{dterm2})  are exactly the ones found in previous QFT analysis \cite{Anber:2025yub} and reproduced earlier in Section \ref{sec:missing}.
The result, $4r +4$, from (\ref{dimensionmoduli}) is precisely the number of moduli for an instanton of topological charge $r \over N$ plus the four moduli associated to the $U(1)$ compensating flux added to facilitate an embedding of the $SU(N)$ fractional instanton in $U(N)$. 

The mystery of the $4r - 4g$ missing moduli (for $r \ne g$)
was pointed out in \cite{Anber:2023sjn} and resolved in \cite{Anber:2025yub} within field theory.
The missing moduli were found in \cite{Anber:2025yub} after a calculation relying on the  results of  \cite{Anber:2023sjn}, which themselves required rather cumbersome and long calculations. The mapping between (\ref{fterm2}, \ref{dterm2}) and the QFT result was described at the end of Section \ref{sec:missing}. Here we observe that the complimentary $D$-brane picture offers a new perspective on the moduli space, leading---in a much quicker fashion---to enumerating the missing moduli. A bonus is that the hyper-K\" ahler structure of the moduli space is automatic, as it is a property of any ${\cal N}=2$ Higgs branch.

\subsection{The enhanced symmetry point} 
\label{sec:enhanced}

As the careful reader will have observed, we derived our expression for the dimension of the moduli space (\ref{dimensionmoduli}), and its local parameterization (\ref{fterm2}, \ref{dterm2}), using the $U(1)^g \times U(1)_\ell$ theory, valid at generic positions of the $g$ ``$k$''-branes and found that   the parameterization of the moduli space in terms of $q_{i}^a, \tilde{q}_{a}^i$ obeying (\ref{fterm2}, \ref{dterm2}) matches the one from the linearized analysis of \cite{Anber:2025yub}. However, when the separation between the $g$ branes (the red and dark blue ones, for $g=2$, on Fig.~\ref{brane5}) vanishes, the worldvolume gauge group enhances to $U(g) \times U(1)_\ell$ as strings connecting the $g$ different  branes in the ``$k$'' stack give rise to extra massless gauge bosons. This does not change the dimension of the Higgs branch and thus the conclusion (\ref{dimensionmoduli}) remains, as we show below (it may bring a moduli space parameterization relevant for the nonlinear analysis of the manifold of self-dual fractional instantons).

To begin with the analysis at the enhanced symmetry point, we first enumerate the matter fields. We use $a=1,...,r/g$ to denote the bifundamental hypermultiplets as before, while $i =1,...,g$ is now an index in the fundamental or antifundamental representation of $U(g)$. 
\begin{center}
\begin{tabular}{c | c | c |c}  \text{field} & $U(1)_\ell$ charge & $U(g)$ representation & multiplicity\\ $q^a_i$ & $1$ & $\Box$ & $a=1,...,{r\over g}$ ($r\over g$ half-hypers) \\ $\tilde{q}_a^i$ & $-1$ & $\overline\Box$ & $a = 1,...,{r\over g}$ ($r\over g$ half-hypers) \\ $\phi_A$& 0 & singlet & $A=1,2$ (hyper) and $A=3$ ($\in$ vector) \\
$\hat\Phi_A$ & 0 & adjoint &$A=1,2$ (hyper) and $A=3$ ($\in$ vector) \end{tabular}
\end{center}
As before, $\phi_{1,2}$ comprise the $U(1)_\ell$ adjoint (i.e. singlet) hypermultiplet, while $\hat\Phi_{A=1,2}$ is the  adjoint hypermultiplet of $U(g)$.\footnote{Its  diagonal components appeared in our  analysis of the $U(1) \times U(1)^g$ theory as $\phi_{1 i}$, $\phi_{2 i}$, uncharged under $U(1)^g$.} The adjoint chiral superfield in the vector multiplet of $U(g)$ is now similarly denoted by $\hat\Phi_3$.
The superpotential (\ref{superpotential1}) becomes upgraded to
\begin{eqnarray}
\label{superpotential2}
W = \text{tr}\; \hat\Phi_3 [\hat\Phi_1, \hat\Phi_2] + \tilde{q}^i_a \; \hat\Phi_{3  i}^{~~ j} \; q_{j}^a + \phi_3 \tilde{q}^i_a q_{i}^a,
\end{eqnarray}
 with all repeated indices summed over. The new coupling, compared to (\ref{superpotential1}), is the coupling between the adjoint in the vectormultiplet $\hat \Phi_3$ and the two adjoint chiral superfields $\hat\Phi_A$ in the adjoint hypermultiplet of $U(g)$. 
 The $q$ and $\tilde{q}$ $F$-term conditions are now
 \begin{eqnarray}\label{fterm3}
 \tilde{q}_a^i (\hat\Phi_{3  i}^{~~ j} + \delta_{i}^j \phi_3) = 0, ~~   (\hat\Phi_{3  i}^{~~ j} + \delta_{i}^j \phi_3) q_j^a = 0, ~i=1,...,g, ~ a = 1,...,{r\over g}~.
 \end{eqnarray}
As before, we work on the Higgs branch, and demanding that all branes be on top of each other in the noncompact directions orthogonal to their worldvolume, we conclude that $\hat \Phi_3 = - \phi_3 {I}_g$, i.e. $\hat\Phi_3$ is proportional to the unit matrix. The  $\hat\Phi_{1,2}$ $F$-term conditions then demand that 
\begin{eqnarray}\label{fterm4}
[\hat\Phi_3, \hat\Phi_A] = 0, ~ A=1,2, 
\end{eqnarray}
which in view of $\hat\Phi_3 \sim I_g$  imposes no constraints on $\hat\Phi_A$. We are then left with the $\hat\Phi_3$ and $\phi_3$ $F$-term conditions:
\begin{eqnarray}\label{fterm5}
[\hat\Phi_1, \hat\Phi_2]_i^{~j} + \sum\limits_{a=1}^{r/g} q_i^a \tilde{q}_a^j = 0, ~\sum\limits_{i=1}^{g} \sum\limits_{a=1}^{r/g} q_i^a \tilde{q}_a^i = 0 ~~ (\text{$g^2$ complex conditions}),
\end{eqnarray}
where, as before, the second, $\phi_3$, condition is redundant.
Finally, the $D$-term conditions from $U(g)$, written after using the completeness relation for the $U(N)$ generators:
\begin{eqnarray}\label{dterm5}
\sum\limits_{a=1}^{r/g} \left(q^{\dagger \; i}_a \; q_j^a - \tilde{q}^{  i}_a \; \tilde{q}_j^{\dagger \; a}\right) + \sum_{A=1}^2\; [\hat\Phi_{A}, \hat\Phi_A^\dagger]^i_j =0 ~~ (\text{$g^2$ real conditions}),
\end{eqnarray} as well as $g^2$ constraints due to modding by $U(g)$; as before, the $U(1)_\ell$ conditions are irrelevant  and hence omitted. 

Counting the number of hypermultiplets, we find $4 g^2 + 4 + 4r $ ($4g^2$ $\hat\Phi_{1,2}$, $4$ $\phi_{1,2}$ and $4 r$ $q_i^a, \tilde{q}^i_a$) real components, while the number of $F$-, $D$-term, and gauge redundancy real constraints is $2 g^2 + g^2  + g^2 = 4 g^2$. Subtracting the two, we find that the Higgs branch dimension is still given by (\ref{dimensionmoduli}). The novelty is that now non-commuting directions in the adjoint hypermultiplets of $U(g)$ may be involved. As already mentioned, the linearized analysis of the moduli space in \cite{Anber:2025yub} agreed with the simpler parameterization of (\ref{fterm2}, \ref{dterm2}), which corresponds to $\hat\Phi_A$ diagonal and hence commuting. In a field theory analysis, the  moduli $\hat\Phi_{1,2}$ with $[\hat\Phi_{1},\hat\Phi_{2}] \ne 0$ from (\ref{fterm5})  would reside in    $U(g) \in U(k) \in SU(N)$. Such contributions are not captured by the linearized analysis of \cite{Anber:2025yub}.

The hyper-K\" ahler structure of the local parameterization of the moduli space follows from the fact that it is a supersymmetric Higgs branch in a theory with eight supercharges (via the ``hyper-K\" ahler quotient'' construction of \cite{Hitchin:1986ea}). We shall not make this explicit, but just refer to the general result. We only note that to make   the $SU(2)_R$ symmetry explicit, one introduces first the $r \over g$ $SU(2)_R$ doublets $\chi^a_i$, which are also $U(g)$ fundamentals
\begin{eqnarray}
\chi^a_i \equiv \left(\begin{array}{c} q_i^a \cr \tilde{q}^{\dagger \; a}_i \end{array}\right),~ i=1,...,g, ~ a=1,..,{r \over g}.
\end{eqnarray}
In terms of $\chi_a$, with $\sigma^C$ the Pauli matrices acting in $SU(2)_R$ space (note that $(\hat\Phi_1, \hat\Phi_2^\dagger)$ is also an $SU(2)_R$ doublet), we have that
 \begin{eqnarray}
 \chi_a^{\dagger \;i} \sigma^C \chi^a_j = \left\{\begin{array}{cc}q^{\dagger \; i}_a \;  \tilde{q}^{\dagger \; a}_j + \tilde{q}_a^i  \; q^a_j &, \; \text{for} \; C=1, \cr  - i q^{\dagger \; i}_a 
 \; \tilde{q}^{\dagger \; a}_j +i\tilde{q}_a^i  \; q^a_j &,\;  \text{for} \;C=2, \cr 
 q^{\dagger \; i}_a \; q^a_j - \tilde{q}_a^i  \; \tilde{q}^{\dagger \; a}_j &, \;\text{for} \; C=3.\end{array}  \right. \end{eqnarray}
 In the first two lines we recognize the sum and difference of the $F$-term contributions of $q, \tilde{q}$ while the third is their contribution to the $D$-term.
 Using the 't Hooft symbols $\eta_{\mu\nu}^C$,\footnote{\label{footnote:etasymbols}For $\mu=1,2,3$, $\eta_{\mu\nu}^C = \epsilon_{C \mu\nu}$, $\eta^C_{4 \mu} = - \delta_{C \mu}$, $\eta^C_{\mu 4} =  \delta_{C \mu}$, while $\eta^C_{44}=0$ and $\bar\eta^C_{\mu\nu} = (-1)^{(\delta_{\mu4} + \delta_{\nu 4})} \eta_{\mu\nu}^C$. } the $F$- and $D$-term conditions for $U(g)$, eqns.~(\ref{fterm5}, \ref{dterm5}), are written as an $SU(2)_R$ triplet
  \begin{eqnarray}\label{hyperk}
\sum\limits_{a=1}^{r \over g}   \chi_a^{\dagger \;i} \sigma^C \chi^a_j -  i \eta_{\mu\nu}^C\;    [X_\mu, X_{\nu}]_j^{\;\; i} &=&0, ~  X_\mu^\dagger = X_\mu, \\
 ~ \text{where}~ \hat\Phi_1 &=& X_1 + i X_2,~ \hat\Phi_2 = X_3 +  i X_4,\nonumber 
  \end{eqnarray}
 and we recall that tracing over the $U(g)$ $i,j$-indices gives the $U(1)$ $F,D$-flatness conditions. Eqn.~(\ref{hyperk}) shows explicitly the $SU(2)_R \times SU(2)_L \simeq SO(4)$ global symmetry of the flatness conditions, with the $SO(4)$ acting on the $\mu$-index of $X_\mu$ as a vector.

{\flushleft{\bf The $r=g$ case:}} The importance of  $r=g$ is that it this is the only case  where, as per the discussion at the end of Section \ref{sec:missing}, we have strong evidence that  the moduli space is fully understood in QFT. Recall also that the $r=g$ case only captures some of the ${1 \over N} \le Q < 1$ instantons, as ${\rm{gcd}}(k,r)$ can equal $r$ only for $r < N$.

For general $g=r$, putting $\chi =0$ (there is only a single $SU(2)_R$ doublet for $g=r$) and taking commuting $X_\mu$ (thus simultaneously diagonalizable) in eqn.~(\ref{hyperk}) reproduces the moduli space found in Section \ref{sec:u1branch}. This, however, leaves open the possibility that there are other solutions of (\ref{hyperk}). To address this, we will now show, for $g=r=2$ that the $\chi=0$ solution with diagonal $X_\mu$ is the only one. The more general $r>2$ case is left for future study.

Thus, we shall now show that for $r=g=2$, the  $U(2) \times U(1)_\ell$ Higgs branch equations yield the same moduli space as (\ref{fterm2}, \ref{dterm2}), described only by the $4 \times 2 + 4$ geometric moduli $\phi_1, \phi_2, \phi_1^{i}, \phi_2^{i}$ ($i=1,...,2$),  already present in (\ref{hatYk}, \ref{hatYell}). Begin by noting that  the $U(1)$ $F$-term condition in (\ref{fterm5}) is obeyed upon using an $SU(2)$ gauge transformation to rotating $q$ to the form given below and then solving the $F$-term condition by taking $\tilde q$ to be orthogonal, thus \begin{eqnarray}\label{show1}
q = \left(\begin{array}{c} v \cr 0 \end{array}\right), ~ \tilde{q} = (0 \; v')~.
\end{eqnarray}
Taking the trace over the $SU(2)$ indices in (\ref{dterm5}) gives the $U(1)_\ell$ $D$-term condition, which then requires that 
\begin{eqnarray}
\label{show2}
|v| = |v'|~.
\end{eqnarray}
Going back to the $SU(2)$ part of (\ref{fterm5}) and substituting (\ref{show1}), we find that 
\begin{eqnarray}\label{show3}
[\hat\Phi_1, \hat\Phi_2]_i^{~j}  + \delta_{i1} \delta^{j2} v v' = 0~.
\end{eqnarray}
The parameterization (\ref{show1}) is invariant under gauge transforms in the Cartan, hence we can make the phase of $v'$ zero, which still leaves an arbitrary phase in $v$.
Next, we rewrite (\ref{show3}) using the Cartan basis of $SU(2)$ generators 
 $\tau^3 = \left( \begin{array}{cc} {1 \over 2} & 0 \cr 0 & -{1 \over 2} \end{array} \right)$,  
 $\tau^{+} =  \left( \begin{array}{cc} 0& 1 \cr 0 & 0 \end{array} \right)$, and $\tau^- = (\tau^+)^\dagger$.  
We expand $\hat\Phi_A = 
\phi_{A}^3 {\tau^3} + \phi_A^+ \tau^+ + \phi_A^- \tau^-$, and, recalling $[\tau^3, \tau^{\pm}] = \pm \tau^\pm$ and $[\tau^+, \tau^-] = 2 \tau^3$, we find
 that $[\hat\Phi_1, \hat\Phi_2] = (\phi_1^3 \phi_2^+ - \phi_1^+ \phi_2^3) \tau_+ -  (\phi_1^3 \phi_2^- - \phi_1^- \phi_2^3) \tau_- + 2 (\phi_1^+ \phi_2^- - \phi_1^- \phi_2^+) \tau^3$. Then the $F$-term equations (\ref{show3}), which we can rewrite as $[\hat\Phi_1, \hat\Phi_2] + vv' \tau^+=0$, require  that
 \begin{eqnarray}\label{show4}
 \phi_1^3 \phi_2^+ - \phi_1^+ \phi_2^3 &=& - vv',\nonumber \\
  \phi_1^3 \phi_2^- &=& \phi_1^- \phi_2^3 , \\ 
  \phi_1^+ \phi_2^- &=& \phi_1^- \phi_2^+ ~.\nonumber
 \end{eqnarray}
The $SU(2)$ $D$-term equations (\ref{dterm5}), now multiplied by $\tau^3$ or $\tau^+$ and traced over,  using (\ref{show1}, \ref{show2}) and the above expansion of $\hat\Phi$ into components, impose the conditions:
 \begin{eqnarray}\label{show5}
0 &=& |v|^2 + \sum_{A=1}^2 \tr \hat\Phi_A^\dagger [\tau^3, \hat\Phi_A] = |v|^2 +   |\phi_1^+|^2+ |\phi_2^+|^2 - |\phi_1^-|^2- |\phi_2^-|^2, \nonumber \\
0 &=& \sum_{A=1}^2 \tr \hat\Phi_A^\dagger [\tau^+, \hat\Phi_A] =  - \phi_1^{ +  *} \phi_1^3 + \phi_1^{3  *} \phi_1^{- \; }  - \phi_2^{ +  *} \phi_2^3 + \phi_2^{3  *} \phi_2^{- \; }, 
 \end{eqnarray}
while the $\tau^-$ $D$-term equation is the complex conjugate of the last. 
 
One solution of  (\ref{show4}) and (\ref{show5})  occurs for $|v|=|v'|=0$. This is given by setting $\phi_{1,2}^\pm = 0$ and leaving undetermined $\phi_1^3$ and $\phi_2^3$. That this solves the above equations follows upon inspection. Most notably, it coincides with the solution of (\ref{fterm2},\ref{dterm2}) and with the field theory analysis (also, it is easy to show that this is the only solution at $v=0$, since the $F$-term equation now requires that $\hat\Phi_A$ commute). 

We now want to show that there is no solution of the $F$- and $D$-term equations for $v \ne 0$. From the first $D$-term equation in (\ref{show5}) we conclude that at least one of $\phi_2^-$ or $\phi_1^-$ have to be nonzero. Assume $\phi_1^-$ has a nonzero component. Then, the second equation  in (\ref{show4}) yields $\phi_2^3 = \phi_1^3 {\phi_2^- \over \phi_1^-}$. Inserting this into the first $F$-term equation in (\ref{show4}), we obtain 
\begin{eqnarray}\label{show6}
\phi_1^3 \phi_2^+ - \phi_1^+ \phi_1^3 {\phi_2^- \over \phi_1^-} = - v v' 
\implies {\phi_1^3 \over \phi_1^-} (\phi_2^+ \phi_1^- - \phi_1^+ \phi_2^-) = - vv'   \implies {\phi_1^3 \over \phi_1^-} \times 0 = - v v', \nonumber \\
\end{eqnarray}
where in the last term we used the third $F$-term equation from (\ref{show4}). The last equation in (\ref{show6}) is inconsistent,\footnote{This can be repeated by assuming, instead, that $\phi_2^-$ is nonzero, arriving at ${\phi_2^3 \over  \phi_2^-} \times 0 = - v v'$ instead.} unless $vv'=0$, showing, by the previous paragraph, that for $r=g$ the moduli space agrees with the one already found in QFT, and involves only the geometric moduli $\phi_{1,2}, \phi_{1,2}^i, i ={1,...,r}$.

   We have not performed a similar study for $r=g$ for general  $r>2$.
 We only sketch how this can be approached.
   As for $r=g$, we only have one $SU(2)_R$ doublet $\chi$, which we write a $2 \times r$ dimensional matrix, with the Pauli matrix in (\ref{hyperk}) acting on each column (i.e.~the $SU(2)_R$ acts on the upper index below):
   \begin{eqnarray}\label{chinew}
\chi =    \left(\begin{array}{cccccc}\chi_1^1 & \chi_2^1& \chi_3^1& ...& \chi_{r-1}^1& \chi_r^1 \cr \chi_1^2 & \chi_2^2& \chi_3^2& ...& \chi_{r-1}^2& \chi_r^2  \end{array}\right) ~\rightarrow~ \chi =     \left(\begin{array}{cccccc}v_1 & 0& 0& ...& 0& 0\cr 0 & v_2 &0 & ...& 0 & 0 \end{array}\right),  v_{1}, v_{2} \in \mathbb{R}, 
   \end{eqnarray}
where the form after the arrow results after the a $U(r) \times SU(2)_R$ rotation. 
  This form implies that the contribution of $\chi$ to (\ref{hyperk}) vanishes if any of the $U(r)$ indices  $i,j$ take values from  $3,...,r$.
  On the other hand in the $i,j=1,2$ subspace, from (\ref{chinew}), the $\chi$ contribution to (\ref{hyperk}) is,
\begin{eqnarray}
  (\chi^{\dagger} \sigma^C \chi)^i_j  = (\sigma^C)^i_j v_j v_i, ~ i,j=1,2,\end{eqnarray}
  with no sum over repeated indices. The $U(r)$ trace condition implies then that $v_1 = v_2 \equiv v \in \mathbb{R}$. Thus, taking all of the above into account (\ref{hyperk})becomes
  \begin{eqnarray}
  \label{flat2}
v^2 (\sigma^C)^i_j  -    i\eta_{\mu\nu}^C\;    [X_\mu, X_{\nu}]_j^{\;\; i} &=&0, ~ \{ i,j \in \{1,2\} \},\nonumber \\
   i\eta_{\mu\nu}^C\;    [X_\mu, X_{\nu}]_j^{\;\; i} &=&0, ~ \{i \in \{3,...,r\} \; \text{or} \;   j \in \{3,...,r\} \},
  \end{eqnarray}
  and we note that for the purpose of studying the moduli space $X_\mu$ can be considered traceless hermitean $SU(r)$ adjoints (the trace part is proportional to the unit matrix and  drops out of (\ref{flat2})). Thus, this can be studied using a Cartan-Weyl basis, similar to what we did for $r=2$. 
 As the general case promises to be quite complicated, we leave this for future work.  
  
 \bigskip
 
 {\flushleft \bf Acknowledgment:} This work is supported by an NSERC Discovery Grant. The author expresses his gratitude to Mohamed Anber for   discussions  and comments on the manuscript and to A.W. Peet for discussions and sharing their insights on $D$-branes.

\bibliography{Nahm3.bib}
  \bibliographystyle{JHEP}

\end{document}